\documentclass[11pt,a4paper]{article}
\usepackage{mathtools}
\usepackage{bbold}
\usepackage[normalem]{ulem}
\usepackage{amssymb}
\usepackage{amsmath}
\usepackage[low-sup]{subdepth}
\usepackage{jheppub}

\newcommand{\p}{\partial}
\newcommand{\dg}{\dagger}
\newcommand{\dd}{\delta}
\newcommand{\bb}{\begin{equation}}
\newcommand{\ee}{\end{equation}}
\newcommand{\ba}{\begin{align}}
\newcommand{\ea}{\end{align}}
\newcommand{\lb}{\left(}
\newcommand{\rb}{\right)}

\DeclareMathOperator{\tr}{tr}
\DeclareMathOperator{\adj}{adj}
\renewcommand{\Im}{\text{Im}}

\newcommand{\one}{\mathbb{1}}
\DeclarePairedDelimiter\abs{\lvert}{\rvert}%
\newcommand{\avg}[1]{{\overline{#1}}}
\newcommand{\ah}[1]{\left[ #1 \right]_\text{ah}}
\newcommand{\hc}{\text{h.c.}}

\title{Implicit schemes for real-time lattice gauge theory}
\author{Andreas Ipp,}
\author{David M\"uller}
\emailAdd{ipp@hep.itp.tuwien.ac.at}
\emailAdd{david.mueller@tuwien.ac.at}
\affiliation{Institut f\"ur Theoretische Physik, Technische Universit\"at Wien, 1040 Vienna, Austria}
\date{\today}

\abstract{
We develop new gauge-covariant implicit numerical schemes for classical real-time lattice gauge theory. A new semi-implicit scheme is used to cure a numerical instability encountered in three-dimensional classical Yang-Mills simulations of heavy-ion collisions by allowing for wave propagation along one lattice direction free of numerical dispersion. We show that the scheme is gauge covariant and that the Gauss constraint is conserved even for large time steps.
}
\arxivnumber{}

\begin{document}
\maketitle
\flushbottom
\newpage
\section{Introduction}

Color Glass Condensate (CGC) effective theory \cite{Iancu:2003xm} applies classical Yang-Mills theory to the area of high energy heavy-ion collisions.
In the CGC description high energy nuclei can be treated as thin sheets of fast moving color charge which generate a classical gluon field. The collision of two such sheets produces the Glasma \cite{Lappi:2006fp}, which behaves classically at the earliest stages of the collision.
Due to classical Yang-Mills theory being non-linear and the non-perturbative nature of the CGC, computer simulations are commonly used to investigate the time evolution of such systems \cite{Krasnitz:1998ns, Lappi:2003bi, Lappi:2011ju, Schenke:2012wb, Schenke:2012fw}. Here, real-time lattice gauge theory provides a numerical treatment of classical Yang-Mills theory while retaining exact gauge invariance on the lattice. To name a few applications besides CGC and Glasma simulations, real-time lattice gauge theory is widely used for non-Abelian plasma simulations \cite{Berges:2008zt, Ipp:2010uy} and hard thermal loop (HTL) simulations \cite{Rebhan:2005re, Rebhan:2008uj, Attems:2012js}, classical statistical simulations of fermion production \cite{Kasper:2014uaa, Gelfand:2016prm}, in studying sphalerons (in electroweak theory) \cite{Moore:1997sn, Moore:1999ww}, for determining the plasmon mass scale in non-Abelian gauge theory \cite{Lappi:2016ato, Lappi:2017ckt} or when studying perturbations on top of a non-Abelian background field \citep{Kurkela:2016mhu}. 

The real-time lattice gauge theory approach is based on the discretization of Yang-Mills fields on a lattice in terms of so-called gauge links, i.e.\ Wilson lines connecting adjacent lattice sites. Using gauge link variables one can discretize the continuum Yang-Mills action in various ways, the simplest of which is the Wilson gauge action \cite{PhysRevD.10.2445}. Varying this action with respect to the link variables one obtains discretized classical field equations, which are of the explicit leapfrog type. Moreover, in addition to the equations of motion, one also obtains the Gauss constraint, which is exactly conserved by the leapfrog scheme, even for finite time steps. Going further, the accuracy of the numerical approximation can be systematically improved by adding higher order terms to the standard Wilson gauge action \cite{Moore:1996wn}.  

In previous publications \cite{Gelfand:2016yho, Ipp:2017lho} we developed lattice Yang-Mills simulations for genuinely three-dimensional heavy-ion collisions in the CGC framework. Unlike the usual boost-invariant approach, we consider collisions of nuclei with thin, but non-vanishing support along the longitudinal direction (the beam axis) and simulate them in the laboratory frame. This has enabled us to study the effects of finite nuclear longitudinal extent (which is inversely proportional to the Lorentz gamma factor $\gamma$) on the rapidity profile of the produced Glasma after the collision. The numerical scheme in these simulations is based on the standard Wilson gauge action with the fields coupled to external color currents. The treatment of these color charges is closely related to the colored particle-in-cell (CPIC) method \cite{Dumitru:2005hj, Strickland:2007}, which is a non-Abelian extension of the particle-in-cell (PIC) method \cite{Verboncoeur2005} commonly used in (Abelian) plasma simulations.

Unfortunately, these simulations suffer from a numerical instability that leads to an artificial increase of total energy if the lattice resolution is too coarse:
even a single nucleus propagating along the beam axis, which should remain static and stable, eventually becomes unstable. Improving the resolution simply postpones the problem at the cost of much higher computational resources. We realize that this instability is due to numerical dispersion on the lattice inherent to the leapfrog scheme, which renders the dispersion relation of plane waves non-linear. As a consequence of numerical dispersion, high frequency plane waves exhibit a phase velocity that is noticeably less than the speed of light on the lattice. The shape of the pulse of color fields is lost over time. At the same time, the color current ``driving" the nucleus forward will not disperse by construction: the point-like color charges making up the current are simply moved from one cell to the next as the simulation progresses. Thus the shape of the current is always kept intact. This mismatch and the resulting instability is therefore related to the numerical Cherenkov instability \citep{GODFREY1974504}, which can occur in (Abelian) particle-in-cell simulations. Notably, simulations of laser wakefield acceleration \citep{PhysRevSTAB.16.021301}, where electric charges moving at relativistic speeds are coupled to discretized electromagnetic fields, suffer from the same type of instability and many numerical schemes have been devised to cure it \citep{GREENWOOD2004665, PhysRevSTAB.8.042001, Godfrey:2014ava, Godfrey:2015sxa}. A particularly simple solution to the problem is the use of semi-implicit schemes to repair the dispersion relation (i.e.\ making it linear) for one direction of propagation \citep{Novokhatski:2012yd}, which is the approach we take in this work.

In this paper we derive an implicit and a semi-implicit scheme for real-time lattice gauge theory by modifying the standard Wilson gauge action. We obtain two new actions that are gauge invariant (in the lattice sense), are of the same order of accuracy as the original action, but yield an implicit or semi-implicit scheme upon variation.
In the case of the semi-implicit scheme setting the lattice spacing and the time step to specific values can fix the dispersion relation along the longitudinal direction and thus suppress the numerical Cherenkov instability.
We also obtain a modified version of the Gauss constraint that is conserved up to (in principle) arbitrary numerical precision under the discrete equations of motion.

We start with a discussion of the main ideas behind the semi-implicit scheme for the two-dimensional wave equation in section \ref{wave_equation} and for Abelian gauge fields on the lattice in section \ref{abelian_fields}. The concepts are then generalized to non-Abelian lattice gauge theory in section \ref{nonabelian_fields}, where we derive both a fully implicit and the semi-implicit scheme. Finally, we verify numerically that the Cherenkov instability can be suppressed using the new scheme and that the Gauss constraint is conserved in section \ref{tests}.

\section{A toy model: the 2D wave equation}\label{wave_equation}

The basic ideas behind the numerical scheme we are after can be most easily explained using a simple toy model, namely the two-dimensional wave equation. We start by giving a few definitions and then derive three different numerical schemes by discretizing the action of the system in different ways and using a discrete variational principle. The schemes obtained through this procedure are known as variational integrators, which exhibit useful numerical properties such as conserving symplectic structure and retaining symmetries of the discrete action \citep{marsden_west_2001, Lew2016}.
We will see how the exact discretization of the action affects the properties of the numerical scheme and in particular how numerical dispersion can be eliminated.

We consider a real-valued scalar field $\phi(x)$ in 2+1 with mostly minuses metric signature $(+1,-1,-1)$ and set the speed of light to $c=1$. The action is given by
\bb \label{eq:wave_action}
S[\phi] = \intop_x \frac{1}{2} \sum_\mu \p_{\mu}\phi\p^{\mu}\phi,
\ee
which upon demanding that the variation of the action vanishes
\bb
\dd S = \intop_x \frac{\dd S[\phi]}{\dd \phi(x)} \dd \phi(x) = 0,
\ee
yields the equations of motion (EOM)
\bb \label{eq:wave_eom}
\p_\mu \p^\mu \phi(x) = \p_0^2 \phi(x) - \sum_i \p_i^2 \phi(x) = 0.
\ee
We use Latin indices $i,j,k,...$ to denote the spatial components, $\p_i^2$ is a shorthand for $\p_i \p_i$ (no sum implied) and $\intop_x = \int dx^0 dx^1 dx^2$. Inserting a plane-wave ansatz 
\bb \label{eq:plane_wave_ansatz}
\phi(x) = \phi_0 \exp(i \sum_\mu k_\mu x^\mu),
\ee
with $\phi_0 \in \mathbb{R}$ and $k^\mu = (\omega, k^1, k^2)^\mu$ into the EOM gives the dispersion relation
\bb
\omega = |k| = \sqrt{(k^1)^2+(k^2)^2}.
\ee
Obviously, the phase velocity $v = \omega / |k| = 1$ is constant, i.e.\ there is no dispersion.

Now let us consider a discretized version of this system by approximating space time as an infinite rectangular lattice with grid spacings $a^\mu$. We refer to $a^0$ as the time step and $a^i$ as the spatial lattice spacings. The field $\phi(x)$ is replaced with field values $\phi_x$ defined at the lattice sites $x$ and derivatives are replaced with finite difference expressions. We define the forward difference
\bb
\p_\mu^F \phi_x \equiv \frac{\phi_{x+\mu} - \phi_{x}}{a^\mu},
\ee
and the backward difference 
\bb
\p_\mu^B \phi_x \equiv \frac{\phi_{x} - \phi_{x-\mu}}{a^\mu},
\ee
where we introduced another shorthand notation: $\phi_{x\pm\mu}$ denotes the field at a neighboring lattice site $x\pm a^\mu \hat{e}_\mu$ (no implicit sum over $\mu$) with the unit vector in the $\mu$ direction $\hat{e}_\mu$. We also define the second order central difference
\bb
\p^2_\mu \phi_x \equiv \p^F_\mu \p^B_\mu \phi_x = \frac{\phi_{x+\mu} + \phi_{x-\mu} - 2 \phi_{x}}{(a^\mu)^2}.
\ee
The forward and backward differences are linear approximations to the first order derivative, while the second order difference is accurate up to second order in the lattice spacing $a$.
Equipped with these definitions we could directly discretize the EOM \eqref{eq:wave_eom}, but this is not the approach we will take. The strategy behind variational integrators is to  first discretize the action \eqref{eq:wave_action} and then demand that the discrete variation vanishes.

\subsection{Leapfrog scheme}
One possible way of discretizing the action is 
\bb \label{eq:wave_action_d1}
S[\phi] = \frac{1}{2} V \sum_x \lb \lb \p^F_0 \phi_x \rb ^2  - \sum_i (\p^F_i \phi_x)^2\right),
\ee
where $\sum_x$ is the sum over all lattice sites and $V = a^0 a^1 a^2$ is the space-time volume of a unit cell. Introducing small variations $\dd \phi_x$ of the discrete field at each point, the discrete variation of this action reads
\begin{align}
\dd S &= V \sum_x \lb \p^F_0 \phi_x \p^F_0 \dd \phi_x - \sum_i \p^F_i \phi_x \p^F_i \dd \phi_x \rb \nonumber \\
&= - V \sum_x \lb  \p^2_0 \phi_x - \sum_i\p^2_i \phi_x \rb \dd \phi_x
\end{align}
which upon setting it to zero yields the discretized EOM
\bb \label{eq:wave_eom_d1}
\p^2_0 \phi_x - \sum_i\p^2_i \phi_x = 0,
\ee
where $\p^2_0$ and $\p^2_i$ are second order finite differences.
Here, we made use of summation by parts, i.e.
\bb
\sum_x \p^F_0 \phi_x \p^F_0 \dd \phi_x = - \sum_x \p^B_0 \p^F_0 \phi_x \dd \phi_x,
\ee
which is the discrete analogue of integration by parts. If the field is known in two consecutive time slices we can explicitly solve the EOM \eqref{eq:wave_eom_d1} for the field values in the next time slice:
\bb\label{eq:wave_leapfrog_explicit}
\phi_{x+0} 	= \sum_i \lb\frac{a^0}{a^i}\rb^2 \lb \phi_{x+i} + \phi_{x-i} - 2\phi_x \rb - \phi_{x-0} + 2\phi_x. 
\ee
In fact, this scheme is identical to the explicit leapfrog scheme\footnote{
	The connection to the leapfrog scheme becomes more apparent if we introduce an approximation of the conjugate momentum
	\bb
	\pi_{x+\frac{0}{2}} \equiv \p^F_0 \phi_x,
	\ee
	which is defined naturally between time slices $x^0$ and $x^0+a^0$ (hence the index ``$x+\frac{0}{2}$" in our notation). The EOM can then be written as
	\begin{align}
	\pi_{x+\frac{0}{2}} &= \sum_i \frac{a^0}{\lb a^i \rb^2} \lb \phi_{x+i} + \phi_{x-i} - 2\phi_x \rb + \pi_{x-\frac{0}{2}}, \\
	\phi_{x+0} &= \phi_x + a^0 \pi_{x+\frac{0}{2}}.
	\end{align}
}, which is accurate up to second order in the time step $a^0$ and spatial lattice spacings $a^i$.
Using the plane-wave ansatz \eqref{eq:plane_wave_ansatz} we find the dispersion relation
\bb \label{eq:wave_disp_d1}
\sin^2 \lb \frac{\omega a^0}{2} \rb = \sum_i \lb \frac{a^0}{a^i} \rb^2 \sin^2 \lb \frac{k^i a^i}{2} \rb,
\ee
which is in general non-linear and only yields real-valued (stable) frequencies $\omega$ for all wave vectors $k$ if the Courant-Friedrichs-Lewy (CFL) condition holds
\bb \label{eq:wave_cfl_d1}
\sum_i \lb \frac{a^0}{a^i} \rb^2 \leq 1.
\ee
The discretization errors of this finite difference scheme result in a non-linear dispersion relation, which is usually referred to as numerical dispersion, since this kind of artificial dispersive behavior of plane waves does not show up in the continuum. If it were possible to set $a^0=a^1=a^2$ (the so-called ``magic time-step") the leapfrog scheme would actually be non-dispersive along the lattice axes, but this choice of the parameters is forbidden by the CFL condition in higher dimensions than $1+1$ and would lead to unstable modes.

\subsection{Implicit scheme}
Let us consider a different discretization: we define a new action
\bb \label{eq:wave_action_d2}
S[\phi] = \frac{1}{2} V \sum_x \lb \lb \p^F_0 \phi_x \rb ^2  - \sum_i \p^F_i \phi_x \p^F_i \overline{\phi}_x \rb,
\ee
where $\overline{\phi}_x$ is the temporally averaged field
\bb 
\overline{\phi}_x \equiv \frac{\phi_{x+0} + \phi_{x-0}}{2} \approx \phi_x +\mathcal{O} \lb (a^0)^2 \rb.
\ee
Note that only one of the spatial finite differences in the squared term is temporally averaged.
Since this action differs from the leapfrog action \eqref{eq:wave_action_d1} only up to an error term quadratic in $a^0$, the numerical scheme derived from this action will have the same accuracy as the leapfrog scheme.

Repeating the steps as before we obtain the discretized EOM
\bb \label{eq:wave_eom_d2}
\p^2_0 \phi_x - \sum_i\p^2_i \overline{\phi}_x = 0.
\ee
This is an implicit scheme, which is more complicated to solve compared to the explicit leapfrog scheme given by eq.\ \eqref{eq:wave_leapfrog_explicit}. Here we have to find the solution to a system of linear equations, which can be accomplished using (for instance) iterative methods.

The dispersion relation for this scheme reads
\bb \label{eq:wave_disp_d2}
\sin^2 \lb \frac{\omega a^0}{2} \rb 	= \lb \sum_i \lb \frac{a^0}{a^i} \rb^2 \sin^2 \lb \frac{k^i a^i}{2} \rb \rb \bigg/ \bigg( 1 + 2 \sum_i \lb \frac{a^0}{a^i} \rb^2 \sin^2 \lb \frac{k^i a^i}{2} \rb \bigg).
\ee
This relation can always be solved for real-valued frequencies $\omega$ and therefore the implicit scheme is unconditionally stable. Unfortunately this does not solve the problem of numerical dispersion either, because there is no choice of lattice parameters that results in a linear dispersion relation. 

We quickly summarize: the first action we considered given by eq.\ \eqref{eq:wave_action_d1} gave us the explicit leapfrog scheme, which is rendered non-dispersive but unstable using the ``magic time-step". The second action, eq.\ \eqref{eq:wave_action_d2}, which we obtained by replacing one of the spatial finite differences with a temporally averaged expression, yields an implicit scheme. This scheme is unconditionally stable, but always dispersive. This suggests that a mixture of both discretizations might solve our problem.

\subsection{Semi-implicit scheme} \label{wave_semi_implicit_scheme}
Finally, we consider the action
\bb \label{eq:wave_action_d3}
S[\phi] = \frac{1}{2} V \sum_x \lb \lb \p^F_0 \phi_x \rb ^2  - \lb \p^F_1 \phi_x \rb^2 -  \p^F_2 \phi_x \p^F_2 \overline{\phi}_x \rb,
\ee
where the derivatives w.r.t.\  $x^1$  are treated like in the leapfrog scheme and the derivatives w.r.t.\ $x^2$ involve a temporally averaged expression as in the implicit scheme.
%
%
The EOM now read
\bb \label{eq:wave_eom_d3}
\p^2_0 \phi_x - \p^2_1 \phi_x - \p^2_2 \overline{\phi}_x = 0.
\ee
We call this numerical scheme semi-implicit, because the finite difference equation contains both explicitly and implicitly treated spatial derivatives.
The dispersion relation associated with eq.\ \eqref{eq:wave_eom_d3} is given by
\bb \label{eq:wave_disp_d3}
\sin^2 \lb \frac{\omega a^0}{2} \rb 	= \lb \sum_i \lb \frac{a^0}{a^i} \rb^2 \sin^2 \lb \frac{k^i a^i}{2} \rb \rb \bigg/ \bigg( 1 + 2\lb \frac{a^0}{a^2} \rb^2 \sin^2 \lb \frac{k^2 a^2}{2} \rb \bigg),
\ee
which is stable if
\bb \label{eq:wave_cfl_d3}
\lb \frac{a^0}{a^1} \rb^2 \leq 1.
\ee
The CFL condition \eqref{eq:wave_cfl_d3} now allows us to set $a^0=a^1$. Looking at the dispersion relation \eqref{eq:wave_disp_d3} we notice that for $k^1 \neq 0$, but $k^2 = 0$ the propagation becomes non-dispersive, i.e.\ $\omega = k^1$. For $k^2 \neq 0$ and $k^1 = 0$ the propagation still exhibits numerical dispersion. The scheme defined by the action \eqref{eq:wave_action_d3} therefore allows for non-dispersive, stable wave propagation along one particular direction on the lattice. This principle also extends to systems with more spatial dimensions, where one treats a preferred direction explicitly and all other spatial directions implicitly.

\subsection{Solution method and numerical tests}

To solve the EOM of the implicit or the semi-implicit scheme one has to solve a linear system of equations. This can be accomplished for instance by inverting a band matrix. Alternatively the equations can also be solved in an iterative manner. Taking the latter approach will be readily applicable to lattice gauge theory.
One example for an iterative method is damped (or relaxed) fixed point iteration: the idea is to first rewrite the EOM \eqref{eq:wave_eom_d3} as a fixed point equation
\bb \label{eq:wave_fixed_point}
\phi_{x+0} = F \left[ \phi \right] = 2 \phi_x - \phi_{x-0} + \lb a^0 \rb^2 \lb  \p^2_1 \phi_x + \p^2_2 \overline{\phi}_x \rb,
\ee
and then, starting with an initial guess $\phi^{(0)}_{x+0}$ from e.g.\ the explicit leapfrog evolution, use the iteration
\bb
\phi^{(n+1)}_{x+0} = \alpha \phi^{(n)}_{x+0} + \lb 1 - \alpha \rb F \left[ \phi^{(n)} \right].
\ee
to obtain a new guess $\phi^{(n+1)}_{x+0}$. Here the real-valued parameter $\alpha$ acts as a damping coefficient. Using fixed point iteration might induce other numerical instabilities not covered by the CFL condition \eqref{eq:wave_cfl_d3}. To analyze this we make the ansatz
\bb
\phi^{(n)}_{x+0} = \lambda^n \varphi_x + \phi^{(\infty)}_{x+0},
\ee
where $\phi^{(\infty)}_{x+0}$ is the true solution to eq.\ \eqref{eq:wave_fixed_point} and $\varphi_x$ represents a time-independent error term. The growth of the error is determined by the modulus of $\lambda$. Employing a Fourier ansatz $\varphi_x = \exp \lb i \sum_i k^i x^i \rb$ yields
\bb
\lambda = -2(1-\alpha) \lb \frac{a^0}{a^2} \rb^{2} \sin^{2} \lb \frac{k^2 a^2}{2} \rb +\alpha,
\ee
which is independent of the $k^1$ component and the corresponding lattice spacing $a^1$.
Requiring convergence for high-$k$ modes, i.e.\ $\abs{\lambda} < 1$ for $k^2=\pm \pi / a^2$, we find
\bb \label{eq:fixed_point_stability}
\frac{2 \dd-1}{2 \dd + 1}<\alpha<1,
\ee
where
\bb
\dd = \lb \frac{a^0}{a^2} \rb^2.
\ee
In $d+1$ dimensions, where we treat the $i=1$ direction explicitly and all others $2 \leq i \leq d$ implicitly, the stability condition is given by eq.\ \eqref{eq:fixed_point_stability} with
\bb
\dd = \sum^d_{i=2} \lb \frac{a^0}{a^i} \rb^2.
\ee
Note that for $\dd < 1/2$ damping might not even be necessary. A similar stability condition can be derived also for the implicit scheme, which by itself (just from the plane wave analysis) is unconditionally stable. It is important to keep in mind that the use of fixed point iteration can introduce new instabilities depending on the lattice spacing and the time step.

\begin{figure}[tbp]
	\centering
	\includegraphics{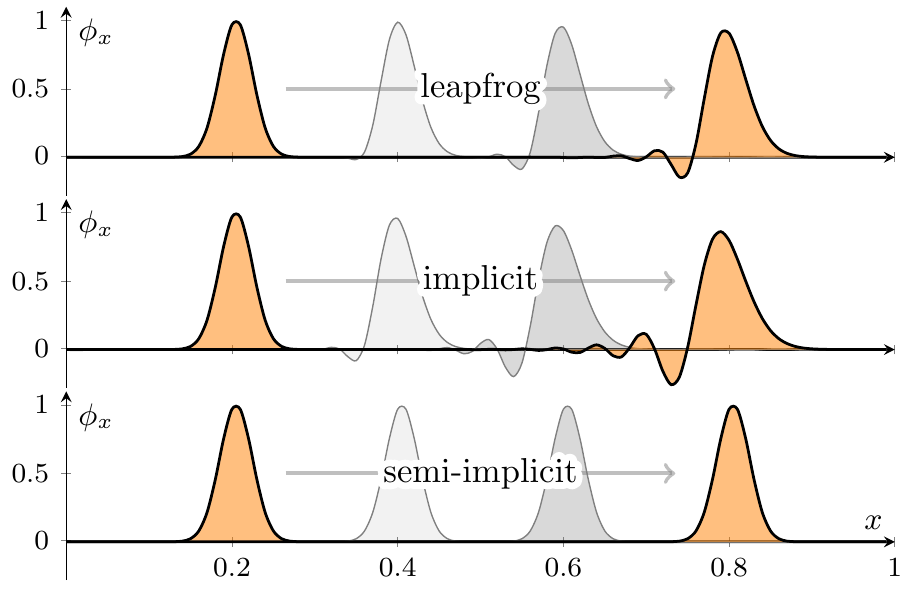}
	\caption{This plot shows a comparison of three different numerical schemes for solving the wave equation in 2+1: (top) explicit leapfrog scheme \eqref{eq:wave_eom_d1}, (middle) the implicit scheme \eqref{eq:wave_eom_d2}, (bottom) semi-implicit scheme  \eqref{eq:wave_eom_d3} with ``magic time-step" $a^0=a^1$. The horizontal axis is the $x^1$ coordinate; the vertical axis is the field amplitude $\phi_x$ in arbitrary units. The $x^2$ coordinate is suppressed. The initial condition (seen on the left) is a Gaussian pulse which propagates to the right under time evolution. Due to numerical dispersion of the leapfrog and implicit scheme the original shape of the Gaussian is lost over time. On the other hand, the dispersion-free semi-implicit solver conserves the original pulse shape. \label{fig:pulse_propagation}}
\end{figure}

Finally, we perform a crucial numerical test: we compare the propagation of a Gaussian pulse using the three different schemes to show the effects of numerical dispersion and in particular that the semi-implicit scheme is dispersion-free. For simulations using the implicit or semi-implicit method we solve the equations using damped fixed point iteration. The results are shown in figure \ref{fig:pulse_propagation}.

The main insight of this section is that the specific discretization of the action completely fixes the numerical scheme of the discrete equations of motion (and the associated stability and dispersion properties) which one obtains from a discrete variational principle.
The use of temporally averaged quantities in the action leads to implicit schemes. If we treat some derivatives explicitly and some implicitly we can end up with a semi-implicit scheme that can be non-dispersive and still stable for propagation along a single direction on the lattice. As it turns out, this is just what we need to suppress the numerical Cherenkov instability we encountered in our heavy-ion collision simulations. In the next section we will see how we can use the same ``trick" for Abelian gauge fields on the lattice.

\section{Abelian gauge fields on the lattice}\label{abelian_fields}
Before tackling the problem of non-Abelian gauge fields on the lattice, it is instructive to see how we can derive a dispersion-free semi-implicit scheme for discretized Abelian gauge fields. We will approach the problem as before: starting with a discretization of the action, we apply a discrete variational principle to derive discrete equations of motion and constraints. Then we will see what modifications to the action are required to obtain implicit and semi-implicit numerical schemes. Since we are dealing with gauge theory we will take care to retain gauge invariance also for the discretized system.

In the continuum the action for Abelian gauge fields reads
\bb
S[A] = -\frac{1}{4} \intop_x \sum_{\mu, \nu} F_{\mu \nu}(x) F^{\mu \nu}(x),
\ee
with the field strength tensor given by
\bb
F_{\mu \nu}(x) = \p_\mu A_\nu(x) - \p_\nu A_\mu(x).
\ee
The field strength and the action are invariant under gauge transformations defined by
\bb
A_\mu(x) \rightarrow A_\mu(x) + \p_\mu \alpha(x),
\ee
where $\alpha(x)$ is an at least twice differentiable function which defines the gauge transformation.

Varying the action with respect to the gauge field leads to
\bb
\dd S[A] = \intop_x \sum_{\nu} \lb \sum_\mu \p_\mu F^{\mu\nu}(x) \rb \dd A_\nu(x)=0.
\ee
The term proportional to $\dd A_0(x)$ leads to the Gauss constraint
\bb
\sum_i \p_i F_{0i}(x)=0,
\ee
while the term proportional to the variation of spatial components gives the EOM
\bb
\p_0 F_{i0}(x) = \sum_j \p_j F_{ij}(x).
\ee
It is trivial to see that the EOM imply the conservation of the Gauss constraint. As we will see next, it is also possible to formulate a discretization of the system where this conservation of the constraint holds exactly.

\subsection{Leapfrog scheme}
We consider a discretized gauge field $A_{x,\mu}$ at the lattice sites $x$. The field strength tensor $F_{x,\mu\nu}$ at $x$ is defined using forward differences
\bb
F_{x,\mu\nu} = \p^F_\mu A_{x,\nu} - \p^F_\nu A_{x,\mu},
\ee
which is antisymmetric in the Lorentz index pair $\mu, \nu$ like its continuum analogue.
Furthermore, the lattice field strength is invariant under lattice gauge transformations given by
\bb \label{eq:lattice_gauge_trans}
A_{x,\mu} \rightarrow A_{x,\mu} + \p^F_\mu \alpha_x,
\ee
where $\alpha_x$ defines the local transformation at each lattice site $x$.
A straightforward discretization of the gauge field action is given by
\bb\label{eq:abelian_action_d1}
S[A] = \frac{1}{2} V \sum_x \lb \sum_i \ F_{x,0i}^2 - \frac{1}{2} \sum_{i,j} F_{x,ij}^2 \rb,
\ee
where $V=\prod_\mu a^\mu$ is the space-time volume of a unit cell. Due to the invariance of $F_{x,\mu\nu}$ under lattice gauge transformations the action is also invariant.

Performing the variation of \eqref{eq:abelian_action_d1} with respect to spatial components $A_{x,i}$ yields the equations of motion. Using the variation of the magnetic part of the action
\begin{align}
	\frac{1}{4} \sum_{x,i,j} \dd \lb F_{x,ij}^2 \rb
    &= \frac{1}{2} \sum_{x,i,j} F_{x,ij} \dd F_{x,ij} \nonumber \\
    &= \frac{1}{2} \sum_{x,i,j} F_{x,ij} \lb \p^F_i \dd A_{x,j} - \p^F_j \dd A_{x,i} \rb \nonumber \\
    &= \sum_{x,i,j} \p^B_j F_{x,ij} \dd A_{x,i},
\end{align}
and the variation of the electric part w.r.t.\ spatial components (denoted by $\dd_s$)
\begin{align}
\frac{1}{2} \sum_{x,i} \dd_s \lb F_{x,0i}^2 \rb
&= \sum_{x,i} F_{x,0i} \p^F_0 \dd A_{x,i} \nonumber \\
&= - \sum_{x,i} \p^B_0 F_{x,i0} \dd A_{x,i},
\end{align}
we find the discrete EOM
\bb \label{eq:abelian_eom_d1}
\p^B_0 F_{x,i0} = \sum_j \p^B_j F_{x,ij},
\ee
which are of the explicit leapfrog type.
We also obtain a discretized version of the Gauss constraint by considering the variation w.r.t.\ temporal components $A_{x,0}$ (denoted by $\dd_t$). With
\begin{align}
\frac{1}{2} \sum_{x,i} \dd_t \lb F_{x,0i}^2 \rb
&= - \sum_{x,i} F_{x,0i} \p^F_i \dd A_{x,0} \nonumber \\
&= \sum_{x,i} \p^B_i F_{x,0i} \dd A_{x,0},
\end{align}
we get the constraint
\bb \label{eq:abelian_constraint_d1}
\sum_i \p^B_i F_{x,0i} = 0.
\ee
Since both the discrete EOM and the constraint follow from the same discretized, gauge-invariant action \eqref{eq:abelian_action_d1}, the Gauss constraint is guaranteed to be automatically conserved under the EOM. We can show this explicitly via
\begin{align}
\p^B_0 \lb \sum_i \p^B_i F_{x,0i} \rb &= -\sum_i \p^B_i \lb \p^B_0 F_{x,i0} \rb \nonumber \\
&= -\sum_{i,j} \p^B_i \p^B_j F_{x,ij}  = 0,
\end{align}
which is equivalent to
\bb
\sum_i \p^B_i F_{x,0i} = \sum_i \p^B_i F_{x-0,0i}.
\ee
This means that if the Gauss constraint is satisfied in one time slice then the EOM will ensure that it remains satisfied in the next time slice.
We can also give a more general proof: consider an infinitesimal gauge transformation
\bb
A'_{x,\mu} = A_{x,\mu} + \p^F_\mu \alpha_x,
\ee
and expand the action $S[A']$ for small $ß\alpha$. We then find
\begin{align}
S[A'] &\simeq S[A] + \sum_{x,y,\mu} \lb \frac{\p S[A']}{\p A'_{x,\mu}} \frac{\p A'_{x,\mu}}{\p \alpha_y} \rb \bigg\rvert_{\alpha=0} \alpha_y +\mathcal{O} \lb \alpha^2 \rb \nonumber \\
&= S[A] + \sum_{x,y,\mu} \frac{\p S[A]}{\p A_{x,\mu}} \p^F_\mu \delta_{xy} \alpha_y +\mathcal{O} \lb \alpha^2 \rb \nonumber \\
&= S[A] - \sum_{x,\mu} \p^B_\mu \frac{\p S[A]}{\p A_{x,\mu}} \alpha_x +\mathcal{O} \lb \alpha^2 \rb.
\end{align}
Since $S[A]$ is invariant for any $\alpha$ it must hold that
\bb
\sum_{\mu} \p^B_\mu \frac{\p S[A]}{\p A_{x,\mu}} = 0,
\ee
or written slightly differently
\bb
\p^B_0 \frac{\p S[A]}{\p A_{x,0}} = - \sum_i \p^B_i \frac{\p S[A]}{\p A_{x,i}}.
\ee
If the equations of motion are satisfied in every time slice, i.e.\ $\frac{\p S[A]}{\p A_{x,i}}=0$, then the Gauss constraint $\frac{\p S[A]}{\p A_{x,0}}$ must be conserved from one slice to the next:
\bb
\p^B_0 \frac{\p S[A]}{\p A_{x,0}} = 0.
\ee
This holds regardless of the exact form of the gauge invariant action $S[A]$. Consequently, it does not matter what kind of discretization of the action we use. As long as $S[A]$ retains lattice gauge invariance in the sense of eq.\ \eqref{eq:lattice_gauge_trans}, we are guaranteed to find that the discrete Gauss constraint is conserved under the discrete equations of motion.

The EOM \eqref{eq:abelian_eom_d1} alone are not enough to uniquely determine the time evolution of the field $A_{x,\mu}$: we must specify a gauge condition. Here we use temporal gauge
\bb
A_{x,0} = 0,
\ee
which we will also use in the case of non-Abelian lattice gauge fields. The EOM \eqref{eq:abelian_eom_d1} then read
\bb \label{eq:abelian_eom2_d1}
- \p^2_0 A_{x,i} = \sum_j \lb \p^B_j \p^F_i A_{x,j} - \p^2_j A_{x,i} \rb.
\ee
Using a plane wave ansatz 
\bb
A_{x,i} = A_i e^{i \lb \omega x^0 - \sum_i k^i x^i \rb},
\ee
with amplitude $A_i$, we find the same non-trivial dispersion relation and CFL stability condition as in the case of the leapfrog scheme for the wave equation in 2+1 (see eq.\ \eqref{eq:wave_disp_d1} and eq.\ \eqref{eq:wave_cfl_d1}). To show this explicitly we first introduce some notation. Taking either a forward (backward) finite difference of the plane wave ansatz yields
\begin{align}
\p^{F}_i A_{x,j} = \frac{e^{-i k^i a^i} - 1}{a^i} A_{x,j}, \\
\p^{B}_i A_{x,j} = \frac{1 - e^{+i k^i a^i}}{a^i} A_{x,j},
\end{align}
which suggests the definition of the forward (backward) lattice momentum
\begin{align}
\kappa^{F}_i = \frac{e^{-i k^i a^i} - 1}{i a^i}, \\
\kappa^{B}_i = \frac{1 - e^{+i k^i a^i}}{i a^i}.
\end{align}
It holds that $\lb \kappa^F_i \rb^\dg = \kappa^B_i$. We define the squared lattice momentum as
\bb
\kappa^2_i \equiv \kappa^{F}_i \kappa^{B}_i = \lb \frac{2}{a^i} \rb^2 \sin^2 \lb \frac{k^i a^i}{2} \rb.
\ee
Furthermore, we can render the lattice momenta dimensionless by multiplying with $a^0$. We define the dimensionless lattice momentum as
\bb\label{eq:dimless_lattice_momentum}
\chi^{F/B}_i = \frac{a^0}{2} \kappa^{F/B}_i,
\ee
and
\bb
\chi^2_i = \chi^F_i \chi^B_i = \lb \frac{a^0}{a^i} \rb^2 \sin^2 \lb \frac{k^i a^i}{2} \rb.
\ee
The factor of $1/2$ in \eqref{eq:dimless_lattice_momentum} is introduced for convenience.
For differences with respect to the time coordinate we find $\chi^2_0 = \sin^2 \lb \omega a^0 / 2 \rb$. 
Using these definitions the Gauss constraint \eqref{eq:abelian_constraint_d1} for the plane wave ansatz can be reduced to
\bb
\sum_i \chi^B_i A_{i} = 0.
\ee
The discrete EOM \eqref{eq:abelian_eom2_d1} can be written as
\bb
\chi^2_0 A_{i} = -\sum_j \lb \chi^B_j \chi^F_i A_{j} + \chi^2_j A_{i} \rb.
\ee
One can eliminate the mixed terms $\chi^B_j \chi^F_i A_{j}$ using the Gauss constraint and find the dispersion relation
\bb
\chi^2_0 = \sum_j \chi^2_j,
\ee
which is equivalent to the dispersion relation of the wave equation eq.\ \eqref{eq:wave_disp_d1}, i.e.\
\bb
\sin^2 \lb \frac{\omega a^0}{2} \rb = \sum_j \lb \frac{a^0}{a^j} \rb^2 \sin^2 \lb \frac{k^j a^j}{2} \rb.
\ee
\subsection{Implicit scheme}
An implicit scheme analogous to the one we derived for the wave equation (see eq.\ \eqref{eq:wave_action_d2}) can be found using the action
\bb\label{eq:abelian_action_d2}
S[A] = \frac{1}{2} V \sum_x \lb \sum_i \ F_{x,0i}^2 - \frac{1}{2} \sum_{i,j} F_{x,ij} M_{x,ij} \rb,
\ee
where we introduce the temporally averaged field-strength
\bb\label{eq:abelian_M_def}
M_{x,ij} = \avg{F}_{x,ij} =  \frac{1}{2} \lb F_{x+0,ij} + F_{x-0,ij} \rb.
\ee
Note that $M_{x,ij}$ differs from $F_{x,ij}$ only by an error term proportional to $\lb a^0 \rb^2$.
Replacing one of the field strengths $F_{x,ij}$ in the quadratic term in the action with its temporally averaged expression $M_{x,ij}$ is analogous to replacing the wave amplitude $\phi_x$ with $\overline{\phi}_x=\frac{1}{2} \lb \phi_{x+0} + \phi_{x-0} \rb$ in the action $\eqref{eq:wave_action_d2}$. The averaged field strength $M_{x,ij}$ is also invariant under lattice gauge transformations:
\bb
M_{x,ij} \rightarrow M_{x,ij} + \p^F_i \p^F_j \avg{\alpha}_x -  \p^F_j \p^F_i \avg{\alpha}_x = M_{x,ij}.
\ee
Varying the action as we did for the leapfrog scheme we find the EOM
\bb \label{eq:abelian_eom_d2}
\p^B_0 F_{x,i0} = \sum_j \p^B_j M_{x,ij},
\ee
and employing temporal gauge we have
\bb \label{eq:abelian_eom2_d2}
- \p^2_0 A_{x,i} = \sum_j \lb \p^B_j \p^F_i \overline{A}_{x,j} - \p^2_j \overline{A}_{x,i} \rb,
\ee
where the fields $A_{x,i}$ have been replaced with temporally averaged expressions $\overline{A}_{x,i}=\frac{1}{2} \lb A_{x+0,i} + A_{x-0,i} \rb$ on the right-hand side. 
The Gauss constraint that arises from varying w.r.t.\ temporal components is simply eq.\ \eqref{eq:abelian_constraint_d1}, because we did not modify the term involving $F_{x,0i}$ or introduce new dependencies on $A_{x,0}$. The discrete EOM \eqref{eq:abelian_eom_d2} still conserve the Gauss constraint due to lattice gauge invariance.
Performing the same steps as in the previous section, we can show that this implicit scheme is unconditionally stable and exhibits the same non-trivial dispersion relation as the implicit scheme for the wave equation, see  eq.\ \eqref{eq:wave_disp_d2}.

\subsection{Semi-implicit scheme} \label{sec:abelian_semi_implicit}

In this section we want to develop a semi-implicit scheme for Abelian gauge fields. Specifically we need an action that allows for dispersion-free propagation of waves in the direction of the $x^1$ coordinate (which we refer to as the longitudinal direction). We call $x^2$ and $x^3$ the transverse coordinates. From now on Latin indices $i, j, k, ...$ refer to transverse indices and the longitudinal index will always be explicit. Our goal is to define the action in such a way that we end up with equations of motion that include explicit differences in the $x^1$ direction, but temporally averaged finite differences in the $x^2$ and $x^3$ direction. This means that we have to modify the $F_{x,i1}^2$ term of the leapfrog action \eqref{eq:abelian_action_d1} such that it results in terms like $\p^F_i A_{x,1} \p^F_1 \avg{A}_{x,1}$. A first guess for a semi-averaged version of field strength $F_{x,i1}$ that could accomplish this is
\bb
\p^F_i \avg{A}_{x,1} - \p^F_1 A_{x,i},
\ee
with $\overline{A}_{x,1} = \frac{1}{2} \lb A_{x+0,1} + A_{x-0,1} \rb$. However, it turns out that such a term is not invariant under lattice gauge transformations \eqref{eq:lattice_gauge_trans}. The problem is that $\avg{A}_{x,1}$ transforms differently than $A_{x,i}$. We have
\begin{align}
\avg{A}_{x,1} \rightarrow \avg{A}_{x,1} + \p^F_1 \avg{\alpha}_x, \\
A_{x,i} \rightarrow A_{x,i} + \p^F_i \alpha_x,
\end{align}
which yields
\bb
\p^F_i \avg{A}_{x,1} - \p^F_1 A_{x,i} \rightarrow \p^F_i \avg{A}_{x,1} - \p^F_1 A_{x,i} + \p^F_i \p^F_1 \lb \avg{\alpha}_x - \alpha_x \rb,
\ee
where the last term $\p^F_i \p^F_1 \lb \avg{\alpha}_x - \alpha_x \rb$ breaks gauge invariance.
To fix this we introduce the ``properly" averaged field strength $\tilde{A}_{x,1}$ given by
\bb \label{eq:abelian_proper_average}
\tilde{A}_{x,1} \equiv \avg{A}_{x,1} - \frac{1}{2} \lb a^0 \rb^2 \p^F_1 \p^B_0 A_{x,0},
\ee
where the last term transforms as
\begin{align}
\frac{1}{2} \lb a^0 \rb^2 \p^F_1 \p^B_0 A_{x,0} &\rightarrow \frac{1}{2} \lb a^0 \rb^2 \p^F_1 \p^B_0 A_{x,0} + \frac{1}{2} \lb a^0 \rb^2 \p^F_1 \p^2_0 \alpha_{x} \nonumber \\
&= \frac{1}{2} \lb a^0 \rb^2 \p^F_1 \p^B_0 A_{x,0} + \p^F_1 \lb \avg{\alpha}_{x} - \alpha_x \rb. 
\end{align}
Here we made use of the exact relation
\bb
\frac{1}{2} \lb a^0 \rb^2 \p^2_0 \alpha_x = \avg{\alpha}_x - \alpha_x.
\ee
Therefore, the transformation property of the properly averaged gauge field $\tilde{A}_{x,1}$ is the same as $A_{x,1}$:
\bb
\tilde{A}_{x,1} \rightarrow \tilde{A}_{x,1} + \p^F_1 \alpha_x.
\ee
It still holds that in the continuum limit the properly averaged gauge field $\tilde{A}_{x,1}$ is the same as $A_{x,1}$ up to an error term quadratic in $a^0$. While the definition \eqref{eq:abelian_proper_average} seems a bit arbitrary at first sight, this way of averaging is more natural using the language of lattice gauge theory: in section \ref{semi_scheme} we will find an intuitive picture in terms of Wilson lines that reduces to \eqref{eq:abelian_proper_average} in the Abelian limit for small lattice spacings.

The properly semi-averaged gauge-invariant field strength is then given by
\bb \label{eq:abelian_W_def}
W_{x,i1} \equiv \p^F_i \tilde{A}_{x,1} - \p^F_1 A_{x,i}.
\ee
In order to keep the field strength explicitly antisymmetric we define $W_{x,1i} = - W_{x,i1}$.
Using these definitions we can guess the action
\bb \label{eq:abelian_action_d3}
S[A] = \frac{1}{2} V \sum_x \bigg(  F_{x,01}^2 + \sum_i F_{x,0i}^2 - \frac{1}{2} \sum_{i,j} F_{x,ij} M_{x,ij}  - \sum_i F_{x,1i} W_{x,1i}\bigg),
\ee 
where the indices $i,j$ denote transverse components. Since we have built the new action from gauge invariant expressions it is also invariant under lattice gauge transformations. Note that the use of $\tilde{A}_{x,i}$ in $W_{x,1i}$ introduces new terms in the action \eqref{eq:abelian_action_d3} dependent on the temporal component of the gauge field. Although these terms disappear in temporal gauge (our preferred choice), they still have an effect on the scheme since we have to perform the variation before choosing a gauge. Therefore we will obtain a modified Gauss constraint compatible with the equations of motion derived from the action \eqref{eq:abelian_action_d3} even after setting $A_{x,0} = 0$.
  
At this point one might ask if the ``proper" averaging procedure has any effect on the implicit scheme of the previous section as well. It turns out that if one defines the averaged field-strength $M_{x,ij}$ of \eqref{eq:abelian_M_def} using the properly averaged gauge field $\tilde{A}_{x,i}$, the action of the implicit scheme (and by extension the EOM and the constraint) remains unchanged. This is due to the fact that the terms proportional to $A_{x,0}$ in \eqref{eq:abelian_proper_average} cancel:
\bb
\p^F_i \tilde{A}_{x,j} - \p^F_j \tilde{A}_{x,i} = \p^F_i \avg{A}_{x,j} - \p^F_j \avg{A}_{x,i}.
\ee
Therefore, no such modification is required in the implicit scheme.

Varying this action w.r.t\ temporal components $A_{x,0}$ yields the modified Gauss constraint
\bb \label{eq:abelian_constraint_d3}
\sum_{i=1}^{d} \p_{i}^{B} F_{x,0i} + \lb \frac{a^{0}}{2} \rb^2 \sum_{i}\p_{i}^{B}\p_{1}^{B}\p_{0}^{F}F_{x,1i} = 0.
\ee
Since $W_{x,1i}$ explicitly depends on $A_{x,0}$, we obtain a correction term to the standard leapfrog Gauss constraint \eqref{eq:abelian_constraint_d1}. The discrete EOM read
\begin{align} \label{eq:abelian_eom_d3_1}
\p^B_0 F_{x,10} &= \frac{1}{2} \sum_i \p^B_i \lb W_{x,1i} + M_{x,1i} \rb,  \\
\p^B_0 F_{x,i0} &= \sum_{j\neq i} \p^B_j M_{x,ij} + \frac{1}{2} \p^B_1 \lb F_{x,i1} + W_{x,i1} \rb. \label{eq:abelian_eom_d3_2}
\end{align}
We have separate EOM for the longitudinal and transverse components of the gauge field. Note that by replacing the averaged expressions $M$ and $W$ with $F$ the EOM reduce to the leapfrog equations as expected.

The propagation of waves in the semi-implicit scheme turns out to be more complicated compared to the leapfrog or implicit scheme: given a wave vector $k$ and a field amplitude $A_{i}$ (such that the Gauss constraint \eqref{eq:abelian_constraint_d3} is satisfied) the dispersion relation becomes polarization dependent, i.e.\ the scheme exhibits birefringence. The amplitude $A_i$ of an arbitrary plane wave
\bb
A_{x,i} = A_i e^{i \lb \omega x^0 - \sum_i k^i x^i \rb},
\ee
has to be split into  a longitudinal and two momentum-dependent transverse components
\bb
\left\{ \vec{A}_{L},\vec{A}_{T,1},\vec{A}_{T,2}\right\} =
\left\{
\left(
\begin{array}{c}
	1\\
	0\\
	0
\end{array}
\right),
\left(
\begin{array}{c}
	0\\
	-\chi_{3}^{B}\\
	\chi_{2}^{B}
\end{array}
\right),
\left(
\begin{array}{c}
	0\\
	\chi_{2}^{F}\\
	\chi_{3}^{F}
\end{array}
\right)\right\}, 
\ee
where $\chi^{F/B}_i$ are dimensionless lattice momenta given by eq.\ \eqref{eq:dimless_lattice_momentum}. In appendix \ref{app_abelian_semi} we find that the components $\vec{A}_{L}$ and $\vec{A}_{T,2}$ have the dispersion relation
\bb \label{eq:semi_dispersion1}
\omega_{1} a^{0}=\arccos\left(\frac{1-\chi_{1}^{2}\left(2+\chi_{2}^{2}+\chi_{3}^{2}\right)}{1+\chi_{2}^{2}\left(2-\chi_{1}^{2}\right)+\chi_{3}^{2}\left(2-\chi_{1}^{2}\right)}\right),
\ee
and the component $\vec{A}_{T,1}$ has a second different dispersion relation
\bb \label{eq:semi_dispersion2}
\omega_{2}a^{0}=\arccos\left(\frac{1-2\chi_{1}^{2}}{1+2\chi_{2}^{2}+2\chi_{3}^{2}}\right).
\ee
Analyzing the stability of the scheme using the two dispersion relations yields that it is stable if
\bb
\chi^2_1 \leq 1,
\ee
which, when requiring stability for all modes, reduces to
\bb
a^0 \leq a^1.
\ee
If we consider the special case of a plane wave with a  purely transverse amplitude and which propagates in the $x^1$ direction (i.e.\ setting the transverse momenta $\chi_2=\chi_3=0$) we find that both dispersion relations agree:
\bb
\omega_1 a^0 = \omega_2 a^0 = \arccos \lb 1- 2 \chi^2_1 \rb.
\ee
This dispersion becomes linear if we set $a^0=a^1$. This explicitly shows that the semi-implicit scheme for Abelian gauge fields allows for dispersion-free, stable propagation along the longitudinal direction if we use the ``magic time-step".
The dispersion relations also agree if we set the longitudinal momentum $\chi_1$ to zero for arbitrary transverse momenta. In general however, wave propagation in this scheme is bifractive.
It would be interesting to see if there are alternative discretizations of the action, which allow for dispersion-free propagation without being bifractive. 

The main result of this section is the action \eqref{eq:abelian_action_d3} which gives rise to the semi-implicit scheme. Here we used a combination of differently averaged field strengths, $M_{x,ij}$ and $W_{x,1i}$, in the action. Our next goal is to generalize these expressions to non-Abelian gauge fields.

\section{Non-Abelian gauge fields on the lattice}\label{nonabelian_fields}
The continuum action for non-Abelian Yang-Mills fields is given by
\bb
S[A] = -\frac{1}{2} \intop_x \sum_{\mu,\nu} \tr \lb F_{\mu\nu}(x) F^{\mu\nu}(x) \rb,
\ee
where the field strength is
\bb
F_{\mu\nu}(x) = \p_\mu A_\nu(x) - \p_\nu A_\mu(x) + i g \left[ A_\mu(x), A_\nu(x) \right].
\ee
The constant $g$ is the Yang-Mills coupling constant and $A_\mu(x) = \sum_a A^a_\mu t^a$ is a non-Abelian gauge field, where $t^a$ are the generators of the gauge group. In the following we use the normalization $\tr \lb t^a t^b \rb = \frac{1}{2} \dd^{ab}$.
Through variation of the action we obtain the Gauss constraint and the equations of motion:
\begin{align}
\sum_i D_i F^{i 0}(x) &= 0, \\
D_0 F^{0 i}(x) &= - \sum_j D_j F^{i j}(x),
\end{align}
where the gauge-covariant derivative acting on an algebra element $\chi$ is given by
\bb
D_\mu \chi(x) \equiv \p_\mu\chi(x) + i g \left[ A_\mu(x),\chi(x)  \right].
\ee
In real-time lattice gauge theory, instead of gauge fields $A_{\mu}(x)$, we use gauge links (or link variables) $U_{x,\mu}$ which are unitary matrices and interpreted as the Wilson lines connecting nearest neighbors on the lattice. $U_{x,\mu}$ is the shortest possible Wilson line on the lattice starting at the site $x$ and ending at $x+\mu$. This is also reflected in the gauge transformations  
\bb \label{eq:lattice_gtr}
U_{x,\mu} \rightarrow V_{x} U_{x,\mu} V^\dg_{x+\mu},
\ee
where $V_x$ is a gauge transformation defined at the lattice site $x$. Gauge links with negative directions are identified as
\bb
U_{x,-\mu} = U^\dg_{x-\mu,\mu}.
\ee
In the continuum limit the gauge links can be approximated by
\bb
U_{x,\mu} \simeq \exp{\lb i g a^\mu A_{x,\mu} \rb},
\ee
where $A_{x,\mu} = \sum_a A^a_{x,\mu} t^a$. We use ``lattice units" for the gauge fields, i.e.\ we absorb a factor of $g a^\mu$ in the definition of the gauge field. The gauge links then read
\bb
U_{x,\mu} \simeq \exp{\lb i \hat{A}_{x,\mu}\rb},
\ee
where $\hat{A}_{x,\mu} \equiv g a^\mu A_{x,\mu}$. For the rest of this paper we drop the hat symbol and just remember to restore factors of $g a^\mu$ whenever necessary. 

\begin{figure}[tbp]
\centering
\includegraphics{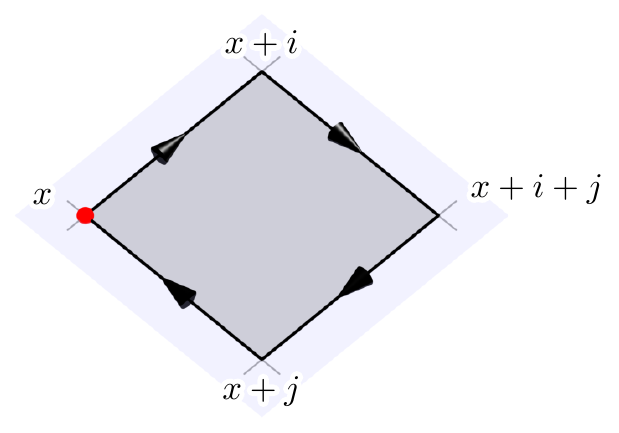} \hspace{25pt}
\includegraphics{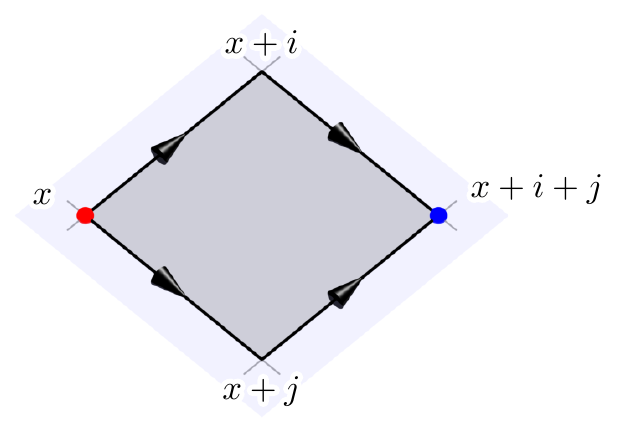}

\caption{ Left: the Wilson line associated with the plaquette $U_{x,ij}$. Right: the Wilson lines associated with the lattice field strength $C_{x,ij}$.
Spatial link variables are drawn as solid black arrows.
While the plaquette starts and ends at the same lattice site $x$ (red dot), the lattice field-strength $C_{x,ij}$ starts at $x$ and ends at $x+i+j$ (blue dot).\label{fig:paths1}}
\end{figure}

The smallest possible Wilson loops that can be constructed on the lattice are the so-called ``plaquettes"
\bb
U_{x,\mu\nu} \equiv U_{x,\mu} U_{x+\mu,\nu} U_{x+\mu+\nu,-\mu} U_{x+\nu,-\nu},
\ee
which can also be written as
\bb
U_{x,\mu\nu} = U_{x,\mu} U_{x+\mu,\nu} U^\dg_{x+\nu,\mu} U^\dg_{x,\nu}.
\ee
The path traced by the plaquette is shown in figure \ref{fig:paths1} on the left. In the continuum limit the plaquettes can be identified with the field strength tensor
\bb
U_{x,\mu\nu} \simeq \exp{\lb i F_{x,\mu\nu} \rb}.
\ee
Here, $F_{x,\mu\nu}$ contains a factor of $g a^\mu a^\nu$. Plaquettes represent a closed Wilson loop and therefore transform locally at the starting (and end) point $x$:
\bb
U_{x,\mu\nu} \rightarrow V_{x} U_{x,\mu\nu} V^\dg_{x}.
\ee
By taking the trace of a plaquette one obtains a gauge invariant expression. In particular it holds that
\begin{align} \label{eq:plaquette_trace}
\tr \lb 2 - U_{x,\mu\nu} - U^\dg_{x,\mu\nu} \rb &\simeq \tr \lb F_{x,\mu\nu}^2 \rb \nonumber \\
&\simeq \frac{1}{2}\sum_a \lb g a^\mu a^\nu F^a_{\mu\nu}(x) \rb^2.
\end{align}
This leads to the standard Wilson gauge action \cite{PhysRevD.10.2445}
\begin{align} \label{eq:wilson_action}
S[U] = \frac{V}{g^2} \sum_x \bigg( & \sum_i \frac{1}{\lb a^0 a^i \rb^2} \tr \lb 2 - U_{x,0i} - U^\dg_{x,0i} \rb \nonumber \\
- \frac{1}{2} & \sum_{i,j} \frac{1}{\lb a^i a^j \rb^2}  \tr \lb 2 - U_{x,ij} - U^\dg_{x,ij} \rb \bigg),
\end{align}
where $\sum_i$ denotes a sum over all spatial components. Using eq.\ \eqref{eq:plaquette_trace} it is clear that the action \eqref{eq:wilson_action} is a discretization of the continuum Yang-Mills action
\bb \label{eq:ym_action}
S[A] = \frac{1}{2} \intop_x \lb \sum_{a,i} F^a_{0i}(x)F^a_{0i}(x)- \frac{1}{2} \sum_{a,i,j} F^a_{ij}(x)F^a_{ij}(x) \rb,
\ee
where we made the split into temporal and spatial components explicit. Since it is built from gauge-invariant expressions, the Wilson gauge action \eqref{eq:wilson_action} is invariant under lattice gauge transformations \eqref{eq:lattice_gtr}.

At this point we remark that the continuum Yang-Mills action \eqref{eq:ym_action} and its discretization \eqref{eq:wilson_action} look very different: while the Yang-Mills action is given in terms of squares of the field strength tensor, the Wilson gauge action is linear in plaquette variables. In terms of plaquettes it is not immediately clear how we might generalize our approach from section \ref{abelian_fields}. Fortunately, the action \eqref{eq:wilson_action} can be written differently so that its functional form is more similar to \eqref{eq:ym_action}. We define (see for instance p.\ 94 of \cite{smit_2002})
\bb \label{eq:lattice_fst}
C_{x,\mu\nu} \equiv U_{x,\mu} U_{x+\mu,\nu} - U_{x,\nu} U_{x+\nu,\mu},
\ee
which transforms non-locally
\bb
C_{x,\mu\nu} \rightarrow V_{x} C_{x,\mu\nu} V^\dg_{x+\mu+\nu}.
\ee
For comparison to the plaquette $U_{x,ij}$, the path traced by $C_{x,ij}$ is shown in figure \ref{fig:paths1} on the right. In the continuum limit $C_{x,\mu\nu}$ can be (up to constant factors) be identified with the field strength $F_{\mu\nu}(x)$: expanding for small lattice spacing we find
\bb
C_{x,\mu\nu} \simeq i g a^\mu a^\nu F_{\mu\nu}(x).
\ee
Most noteworthy is the exact relation
\bb
C_{x,\mu\nu} C^\dg_{x,\mu\nu} = 2 - U_{x,\mu\nu} -U^\dg_{x,\mu\nu},
\ee
with which we can identically rewrite the action as
\bb \label{eq:wilson_action2}
S[U] = \frac{V}{g^2} \sum_x \bigg( \sum_i \frac{1}{\lb a^0 a^i \rb^2} \tr \lb C_{x,0i} C^\dg_{x,0i} \rb -\frac{1}{2} \sum_{i,j} \frac{1}{\lb a^i a^j \rb^2}  \tr \lb C_{x,ij} C^\dg_{x,ij} \rb \bigg).
\ee
This functional form of the rewritten action \eqref{eq:wilson_action2} is now virtually the same as the continuum case \eqref{eq:ym_action}. We will exploit this when generalizing the implicit \eqref{eq:abelian_action_d2} and semi-implicit schemes \eqref{eq:abelian_action_d3} to non-Abelian gauge fields.

Performing the variation of \eqref{eq:wilson_action} or \eqref{eq:wilson_action2} is a bit more involved compared to the wave equation or Abelian gauge fields on the lattice. The degrees of freedom, the gauge links $U_{x,\mu}$, are not just completely general complex matrices but elements of SU(N), i.e.\ unitary matrices with determinant 1:
\begin{align}
U_{x,\mu} U^\dg_{x,\mu} = \one, \\
\det U_{x,\mu} = 1.
\end{align}
This means that one has to perform a constrained variation of the degrees of freedom, which can either be done rather tediously by including the constraints explicitly in the action using Lagrangian multipliers or with the help of a variation that conserves the unitary and determinant constraint. An appropriate variation is given by
\bb \label{eq:link_variation}
\dd U_{x,\mu} = i \dd A_{x,\mu} U_{x,\mu}, 
\ee
where $\dd A_{x,\mu}=\sum_a \dd A^a_{x,\mu} t^a$ is an element of the Lie algebra $\mathfrak{su}$(N), i.e.\ hermitian and traceless. An extended discussion can be found in the appendix \ref{app_var}.

\subsection{Leapfrog scheme}

Following the same procedure as in the case of Abelian gauge fields, we vary w.r.t.\ spatial components $U_{x,i}$ to obtain the discrete EOM and w.r.t.\ temporal components $U_{x,0}$ to find the Gauss constraint.

Starting with the constraint we get (see appendix \ref{app_leapfrog_gauss} for details)
\bb \label{eq:leapfrog_gauss}
\sum_i \frac{1}{\lb a^0 a^i \rb^2} P^a \lb U_{x,0i} + U_{x,0-i}\rb = 0.
\ee
Here we introduced the shorthand $P^a\lb \dots \rb$ to denote
\bb
P^a\lb U \rb = 2 \Im \tr \lb t^a U \rb.
\ee
Varying the spatial link variables we obtain the EOM (see appendix \ref{app_leapfrog_eom} for more details)
%
\bb\label{eq:leapfrog_eom1}
 \frac{1}{\lb a^0 a^i \rb^2} P^a \lb U_{x,i0} + U_{x,i-0}\rb =
- \sum_j \frac{1}{\lb a^i a^j \rb^2} P^a \lb U_{x,i} \lb U_{x+i,j} C^\dg_{x,ij} + C^\dg_{x-j,ji} U_{x-j,j}\rb\rb,
\ee
%
which upon using the definition of $C_{x,ij}$ can be written in the more familiar form
\bb \label{eq:leapfrog_eom}
\frac{1}{\lb a^0 a^i \rb^2} P^a \lb U_{x,i0} + U_{x,i-0}\rb =
\sum_j \frac{1}{\lb a^i a^j \rb^2} P^a \lb U_{x,ij} + U_{x,i-j}\rb.
\ee
As before, time evolution under the EOM conserves the Gauss constraint exactly (see appendix \ref{app_gauss}). In order to actually solve the equations we specify the temporal gauge
\bb
U_{x,0} = \one,
\ee
which enables us to compute the spatial link variables of the next time-slice using past links and the temporal plaquette:
\bb \label{eq:link_evolve}
U_{x+0,i} = U_{x,0i} U_{x,i}.
\ee
The temporal plaquette $U_{x,0i}$ has to be determined from eq.\ \eqref{eq:leapfrog_eom}. For SU(2) this can be done explicitly \citep{Lappi:2016ato,Lappi:2017ckt}: We use the parametrization
\bb 
U_{x,i0} = u^0 \one + i \sum_a u^a \sigma^a,
\ee
where $\sigma^a$, $a\in \{1,2,3\}$ are the Pauli matrices. The real-valued parameters $u^0$, $u^a$ fulfill the constraint 
\bb \label{eq:su2_constraint}
(u^0)^2 + \sum_a u^a u^a = 1.
\ee
The EOM \eqref{eq:leapfrog_eom} can then be written as
\bb \label{eq:su2_evolve1}
\frac{1}{2} u^a = -P^a\lb U_{x,i-0} \rb + \sum_j \lb \frac{a^0}{a^j} \rb^2 P^a \lb U_{x,ij} + U_{x,i-j}\rb.
\ee
To obtain $u^0$ we take the positive branch of eq.\ \eqref{eq:su2_constraint}
\bb \label{eq:su2_evolve2}
u^0=\sqrt{1 - \sum_a u^a u^a},
\ee
assuming that the changes from one time slice to the next are ``small", i.e.\ the temporal plaquette $U_{x,0i}$ will be ``closer" to $\one$ than to $-\one$. 

\subsection{Implicit scheme} \label{implicit_scheme}

Guided by what we learned from the Abelian case in section \ref{abelian_fields}, we would like to replace one of the $C_{x,ij}$ expressions in the Wilson action \eqref{eq:wilson_action2} with a time-averaged equivalent. At the same time we need to retain the gauge invariance of the action. Simply using the temporally averaged expression
\bb \label{eq:C_simple_avg}
\frac{1}{2} \lb C_{x+0,ij} + C_{x-0,ij} \rb,
\ee
is not enough because $C_{x+0,ij}$ and $C_{x-0,ij}$ transform differently. A solution is to include temporal gauge links in order to ``pull back" $C_{x+0,ij}$ and $C_{x-0,ij}$ to the lattice site $x$. This leads us to the definition of the ``properly" averaged field strength
\bb \label{eq:implicit_M_def}
M_{x,ij} \equiv \frac{1}{2} \lb U_{x,0} C_{x+0,ij} U_{x+i+j,-0} + U_{x,-0} C_{x-0,ij} U_{x+i+j-0,0} \rb,
\ee
which transforms like $C_{x,ij}$, i.e.
\bb
M_{x,ij} \rightarrow V_{x} M_{x,ij} V^\dg_{x+i+j}.
\ee
This gauge-covariant averaging procedure can be generalized: consider an object $\mathcal{X}_{x,y}$ that transforms like
\bb \label{eq:X_transf}
\mathcal{X}_{x,y} \rightarrow V_x \mathcal{X}_{x,y} V^\dg_y.
\ee
As an example, $\mathcal{X}_{x,y}$ could be a Wilson line connecting points $x$ and $y$ along some arbitrary path. A time-averaged version of $\mathcal{X}_{x,y}$ is given by
\bb \label{eq:time_avg}
\avg{\mathcal{X}}_{x,y} = \frac{1}{2} \lb U_{x,0} \mathcal{X}_{x+0,y+0} U_{y+0,-0} 
								+ U_{x,-0} \mathcal{X}_{x-0,y-0} U_{y-0,0}\rb,
\ee
where $\mathcal{X}_{x\pm 0,y\pm 0}$ is simply $\mathcal X _{x,y}$ shifted up (or down) by one time step.
It still transforms like eq.\ \eqref{eq:X_transf}, i.e.
\bb
\avg{\mathcal{X}}_{x,y} \rightarrow V_x \avg{\mathcal{X}}_{x,y} V^\dg_y.
\ee
Using this we can write
\bb
M_{x,ij} = \avg{C}_{x,ij}.
\ee
Condensing our notation even further we write
\begin{align}
C^{(+0)}_{x,ij} &= U_{x,0} C_{x+0,ij} U_{x+i+j,-0}, \\
C^{(-0)}_{x,ij} &= U_{x,-0} C_{x-0,ij} U_{x+i+j-0,0},
\end{align}
and
\bb
M_{x,ij} = \frac{1}{2} \lb C^{(+0)}_{x,ij} + C^{(-0)}_{x,ij} \rb.
\ee
It still holds that
\bb
M_{x,ij} \simeq C_{x,ij} + \mathcal{O} \lb \lb a^0 \rb^2 \rb,
\ee
so the use of $M_{x,ij}$ instead of $C_{x,ij}$ does not change the accuracy of the scheme.

Note that temporal gauge renders all temporal link variables trivial and eq.\ \eqref{eq:time_avg} reduces to the simple time average.

%
\begin{figure}[tbp]
\centering
\includegraphics{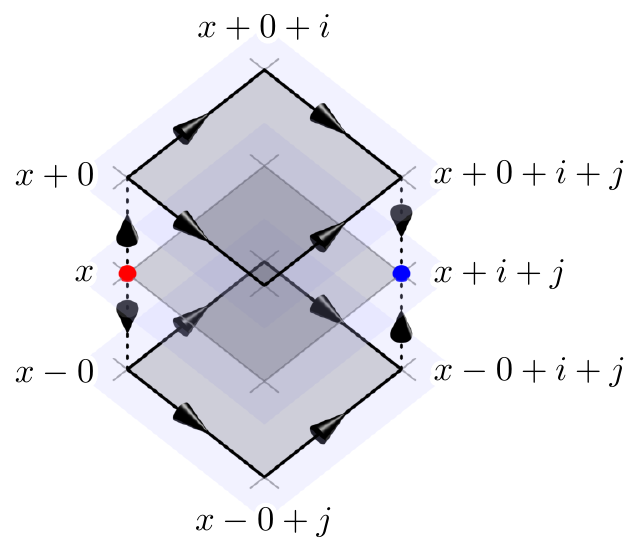} \hspace{25pt}
\includegraphics{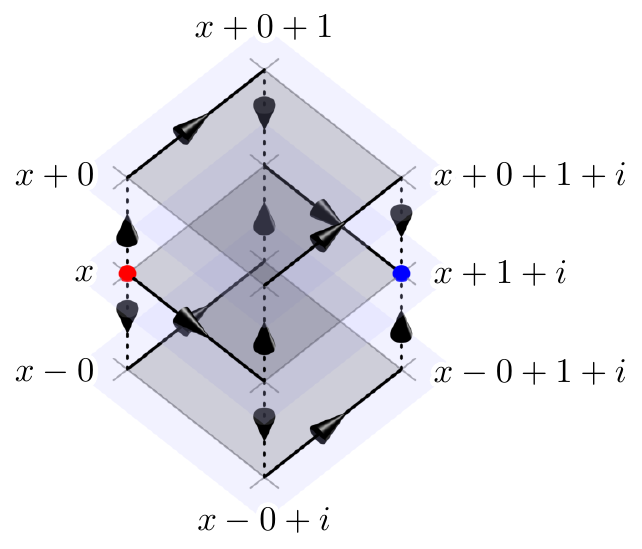}
\caption{Left: the Wilson lines associated with the properly time-averaged field strength $M_{x,ij}$ used in the implicit scheme.
Right: the Wilson lines associated with the partially averaged field-strength $W_{x,1i}$ used in the semi-implicit scheme.
Spatial link variables are drawn as solid black arrows; temporal links as dashed arrows. The shaded planes represent (equal) time slices, i.e.\ the spatial lattice (only two dimensions shown) in subsequent time steps.
The averaged field-strength $M_{x,ij}$ has temporal link connections only at the start and end points. Therefore the diagonal from $x$ to $x+i+j$ is a preferred direction. This asymmetry can be repaired by including $M_{x,i-j}$ terms in the action \eqref{eq:implicit_action}.
We can also see that in the continuous time limit (the vertically stacked time slices would merge into one) the paths traced by $M_{x,ij}$ and $W_{x,1i}$ would become identical to $C_{x,ij}$.
\label{fig:paths2}}
\end{figure}

The inclusion of temporal links in $M_{x,ij}$ breaks a symmetry on the lattice: the diagonal from $x$ to $x+i+j$ is now a preferred direction, which can be seen in figure \ref{fig:paths2}. The loss of symmetry can be mitigated by also including terms like $M_{x,i-j}$ in the action. Therefore we propose the action

%
\bb \label{eq:implicit_action}
S[U] = \frac{V}{g^2} \sum_x \bigg( \sum_i \frac{1}{\lb a^0 a^i \rb^2} \tr \lb C_{x,0i} C^\dg_{x,0i} \rb -\frac{1}{4} \sum_{i, \abs{j}} \frac{1}{\lb a^i a^j \rb^2}  \tr \big( C_{x,ij} M^\dg_{x,ij} \big) \bigg),
\ee
%
where we explicitly include terms with negative spatial indices using the sum $\sum_{\abs{j}}$ over positive and negative components $j$ to keep the action as symmetric as possible. The action is also invariant under time reversal, real-valued (see appendix \ref{app_implicit} for a proof) and gauge invariant. While we have not made any changes to the terms involving temporal plaquettes the spatial plaquette terms now include temporal links and therefore we will also obtain a modified Gauss constraint like in the semi-implicit scheme for Abelian fields.

Performing the variation the same way we did for the leapfrog scheme, we obtain the Gauss constraint (see appendix \ref{app_implicit_gauss})
\bb \label{eq:implicit_gauss}
\sum_{i} \frac{1}{\left(a^{0}a^{i}\right)^{2}} P^a \lb U_{x,0i}+U_{x,0-i} \rb =
- \sum_{\abs i,\abs j} \frac{1}{8} \frac{1}{\lb a^{i}a^{j} \rb^{2}} P^a \lb C_{x,ij}^{(+0)}C_{x,ij}^\dg \rb,
\ee
where $\sum_{\abs i}$ denotes the sum over positive and negative indices $i$. The left hand side (LHS) of \eqref{eq:implicit_gauss} is the same as in the leapfrog scheme eq.\ \eqref{eq:leapfrog_gauss}, but now there is also a new term on the right hand side (RHS) from varying the spatial part of the action. Performing the continuum limit for the Gauss constraint (after multiplying both sides with $a^0$), the RHS term vanishes as $\mathcal{O} \lb \lb a^0 \rb^2 \rb$. This shows that the RHS is not a physical contribution, but rather an artifact of the implicit scheme. Note that the correct continuum limit of the constraint (and the EOM) is already guaranteed by the action.

In a similar fashion as before we perform the variation w.r.t.\ spatial link variables to get the discrete EOM. We find (see appendix \ref{app_implicit_eom})
\bb\label{eq:implicit_eom}
 \frac{1}{\lb a^0 a^i \rb^2} P^a \lb U_{x,i0} + U_{x,i-0}\rb =
- \frac{1}{2} \sum_{\abs j} \frac{1}{\lb a^i a^j \rb^2} P^a \lb U_{x,i} \lb U_{x+i,j} M^\dg_{x,ij} + M^\dg_{x-j,ji} U_{x-j,j}\rb\rb,
\ee
which is formally similar to the leapfrog scheme \eqref{eq:leapfrog_eom1} with $M_{x,ij}$ in place of $C_{x,ij}$, a sum over positive and negative components $j$ (instead of just positive indices) and an additional factor of $1/2$ to avoid overcounting. As a simple check one can replace $M_{x,ij}$ with $C_{x,ij}$ in eq.\ \eqref{eq:implicit_eom} (only introducing an irrelevant error term quadratic in $a^0$) and recover eq.\ \eqref{eq:leapfrog_eom1}.

Compared to the leapfrog scheme, solving eq.\ \eqref{eq:implicit_eom} is more complicated: it is not possible to explicitly solve for the temporal plaquette $U_{x,i0}$ anymore because $M_{x,ij}$ on the RHS involves contributions from both ``past" and ``future" link variables. Completely analogous to the case of Abelian gauge fields on a lattice, we obtain an implicit scheme by introducing time-averaged field strength terms in the action.

Similar to what we did in section \ref{wave_semi_implicit_scheme}, we propose to solve eq.\ \eqref{eq:implicit_eom} iteratively using damped fixed point iteration. Starting from an initial guess for the future link variable $U^{(0)}_{x+0,i}$, for instance by performing a single evolution step using the leapfrog scheme, we iterate
\begin{align}\label{eq:implicit_eom_iteration1}
 \frac{1}{\lb a^0 a^i \rb^2} \lb \mathcal{U}^a +  P^a \lb U_{x,i-0}\rb \rb &=
- \frac{1}{2} \sum_{\abs j} \frac{1}{\lb a^i a^j \rb^2} P^a \lb U_{x,i} \lb U_{x+i,j} M^{(n)\dg}_{x,ij} + M^{(n)\dg}_{x-j,ji} U_{x-j,j}\rb\rb, \\ \label{eq:implicit_eom_iteration2}
P^a \lb U^{(n+1)}_{x,i0} \rb &= \alpha P^a \lb  U^{(n)}_{x,i0} \rb + \lb 1 - \alpha \rb \mathcal U^{a},
\end{align}
using $U^{(n)}_{x+0,i}$ in $M^{(n)}_{x,ij}$ from the last iteration step to determine  $U^{(n+1)}_{x,i0}$. We first solve eq.\ \eqref{eq:implicit_eom_iteration1} for $\mathcal U^{a}$ and then update $P^a \lb U^{(n+1)}_{x,i0} \rb$ using eq.\ \eqref{eq:implicit_eom_iteration2}. The parameter $\alpha$ is used as a damping coefficient to mitigate numerical instabilities induced by fixed point iteration. For SU(2) we construct the temporal plaquette $U^{(n+1)}_{x,i0}$ from $P^a \lb U^{(n+1)}_{x,i0} \rb$ using eqs.\ \eqref{eq:su2_evolve1} and \eqref{eq:su2_evolve2} and using temporal gauge we update the link variables via
\bb
U^{(n+1)}_{x+0,i} = U^{(n+1)}_{x,0i} U_{x,i}.
\ee
Then we can repeat the iteration and keep iterating until convergence.

This iteration scheme can be used to solve the EOM \eqref{eq:implicit_eom} until the Gauss constraint \eqref{eq:implicit_gauss} is satisfied up to the desired numerical accuracy. Conversely, this means that unlike the leapfrog scheme, where the Gauss constraint \eqref{eq:leapfrog_gauss} is always satisfied up to machine precision in a single evolution step, the implicit scheme, solved via an iterative scheme, only approximately conserves the Gauss constraint \eqref{eq:implicit_gauss}.
However, in section \ref{tests} we will show that using a high number of iterations the constraint can be indeed fulfilled to arbitrary accuracy. In practice however we find that a lower number of iteration is sufficient for stable and acceptably accurate simulations at the cost of small violations of the constraint. 

It is also immediately obvious that solving the implicit scheme requires higher computational effort compared to the leapfrog scheme. Considering that one has to use the leapfrog scheme for a single evolution step once (as an initial guess) and then use fixed point iteration, where every step is at least as computationally demanding as single leapfrog step, it becomes clear that the use of an implicit scheme is only viable if increased stability allows one to use coarser lattices while maintaining accurate results.

\subsection{Semi-implicit scheme} \label{semi_scheme}

Using our knowledge from sections \ref{abelian_fields} and \ref{implicit_scheme} we can now generalize the semi-implicit scheme to real-time lattice gauge theory. An appropriate generalization of the semi-averaged field strength \eqref{eq:abelian_W_def} is given by
\bb
W_{x,1i} = \frac{1}{2} \lb U^{(+0)}_{x,1} +  U^{(-0)}_{x,1}  \rb U_{x+1,i} - \frac{1}{2} U_{x,i} \lb U^{(+0)}_{x+i,1} +  U^{(-0)}_{x+i,1} \rb, 
\ee
where
\begin{align}
U^{(+0)}_{x,\mu} &= U_{x,0} U_{x+0,\mu} U_{x+\mu+0, -0}, \\
U^{(-0)}_{x,\mu} &= U_{x,-0} U_{x-0,\mu} U_{x+\mu-0, 0}.
\end{align}
We also define $W_{x,i1} \equiv - W_{x,1i}$.
Using the time-averaging notation (see eq.\ \eqref{eq:time_avg}) this can be written more compactly as
\bb
W_{x,1i} = \avg{U}_{x,1} U_{x+1,i} - U_{x,i} \avg{U}_{x+i,1},
\ee
where
\bb \label{eq:nonabelian_proper_average}
\avg{U}_{x,1} \equiv \frac{1}{2} \lb U^{(+0)}_{x,1} +  U^{(-0)}_{x,1} \rb.
\ee
Note that
\bb
\avg{U}_{x,1} \simeq U_{x,1} + \mathcal O \lb \lb a^0 \rb^2 \rb,
\ee
which shows that the semi-averaged field strength $W_{x,1i}$ only differs from $C_{x,1i}$ by an irrelevant error term. Taking the Abelian limit (i.e.\ neglecting commutator terms) of eq.\ \eqref{eq:nonabelian_proper_average} and expanding for small lattice spacing yields
\bb
\avg{U}_{x,1} = 1+ia^{1}\left(\avg{A}_{x,1}-\frac{1}{2}\left(a^{0}\right)^{2}\p_{1}^{F}\p_{0}^{B}A_{x,0}\right)+\mathcal{O}\left(\lb a^1 \rb^{2}\right).
\ee
We find that linear term of the gauge-covariant average \eqref{eq:nonabelian_proper_average} agrees with the expression we constructed in the Abelian semi-implicit scheme \eqref{eq:abelian_proper_average}. While we had to include an ``arbitrary" correction term in the Abelian scheme to fix gauge invariance, the link formalism of lattice gauge theory forces us to only consider closed paths constructed from gauge links in the action and thus naturally leads us to the ``proper" averaging procedure. 

The Wilson line path traced by $W_{x,1i}$, as compared to $M_{x,ij}$, is shown in figure \ref{fig:paths2} on the right. As before, we keep the action as symmetric as possible by also including terms with negative transverse directions, i.e.\ $W_{x,1-i}$. Inspired by the Abelian semi-implicit case \eqref{eq:abelian_action_d3}, we define the new action as
%
\begin{align}
\label{eq:semiimplicit_action}
S[U] = \frac{V}{g^2} \sum_x \bigg(
& \frac{1}{\lb a^0 a^1 \rb^2} \tr \lb C_{x,01} C^\dg_{x,01} \rb + \sum_i \frac{1}{\lb a^0 a^i \rb^2} \tr \lb C_{x,0i} C^\dg_{x,0i} \rb  \nonumber \\
-\frac{1}{4} \sum_{i,\abs j} & \frac{1}{\lb a^i a^j \rb^2}  \tr \lb C_{x,ij} M^\dg_{x,ij} \rb
-\frac{1}{4} \sum_{\abs j} \frac{1}{\lb a^1 a^j \rb^2}  \tr \lb C_{x,1j} W^\dg_{x,1j} + \hc \rb
\bigg),
\end{align}
%
where the sum over $i$ and $j$ only run over transverse coordinates and $x^1$ is the longitudinal coordinate. The purely transverse part of the action uses the same terms as the implicit scheme, see eq.\ \eqref{eq:implicit_action}. The longitudinal-transverse part is now given in terms of $C_{x,1j}$ and $W_{x,1j}$ analogous to eq.\ \eqref{eq:abelian_action_d2}. We have to explicitly include the hermitian conjugate in order to keep the action real-valued.

Varying with respect to temporal components yields the Gauss constraint (see appendix \ref{app_semi_gauss})
\begin{align} \label{eq:semi_gauss}
\sum^3_{i=1} \frac{1}{\left(a^{0}a^{i}\right)^{2}} P^a \lb U_{x,0i}+U_{x,0-i} \rb = - \sum_{\abs i,\abs j} \frac{1}{8} \frac{1}{\lb a^{i}a^{j} \rb^{2}} P^a \lb C_{x,ij}^{(+0)}C_{x,ij}^\dg \rb \nonumber \\
\qquad \qquad - \frac{1}{8} \frac{1}{\lb a^1 \rb^2} P^a \bigg(U^{(+0)}_{x,1} T^\dg_{x,1} + T^{(+0)}_{x,1} U^\dg_{x,1} + U^{(+0)}_{x-1,1} T^\dg_{x-1,1} + T^{(+0)}_{x-1,1} U^\dg_{x-1,1}  \bigg),
\end{align}
where we use the shorthand
\begin{align}
T_{x,1} &\equiv \sum_{ \abs j} \frac{1}{\lb a^j \rb^2} \lb C_{x,1j} U_{x+1+j,-j} - U_{x,-j} C_{x-j,1j} \rb \nonumber \\
& = \sum_{ \abs j}  \frac{1}{\lb a^j \rb^2} \lb 2 - U_{x,j1} -U_{x,-j1} \rb U_{x,1}.
\end{align}
Varying w.r.t.\ spatial links the discrete semi-implicit EOM read (see appendix \ref{app_semi_eom})
\begin{align}\label{eq:semi_eom_1}
 \frac{1}{\lb a^0 a^1 \rb^2} P^a \lb U_{x,10} + U_{x,1-0}\rb =&
- \frac{1}{4} \sum_{ \abs j} \frac{1}{\lb a^1 a^j \rb^2} P^a \bigg( U_{x,1} \bigg( U_{x+1,j} W^\dg_{x,1j} + W^\dg_{x-j,j1} U_{x-j,j} \nonumber \\
& \qquad\qquad\qquad+ \avg{ \lb U_{x+1,j} C^\dg_{x,1j} + C^\dg_{x-j,j1} U_{x-j,j}\rb}  \bigg) \bigg),
\end{align}
and for transverse components
\begin{align}\label{eq:semi_eom_i}
 \frac{1}{\lb a^0 a^i \rb^2} P^a \lb U_{x,i0} + U_{x,i-0}\rb =&
- \frac{1}{2} \sum_{\abs j} \frac{1}{\lb a^i a^j \rb^2} P^a \lb U_{x,i} \lb U_{x+i,j} M^\dg_{x,ij} + M^\dg_{x-j,ji} U_{x-j,j} \rb \rb \nonumber \\
&- \frac{1}{4} \frac{1}{\lb a^i a^1 \rb^2} \sum_{\abs 1} P^a \bigg( U_{x,i} \bigg( \lb U_{x+i,1} W^\dg_{x,i1} + W^\dg_{x-1,1i} U_{x-1,1} \rb \nonumber \\
& + \lb \avg{U}_{x+i,1} C^\dg_{x,i1} + C^\dg_{x-1,1i} \avg{U}_{x-1,1} \rb \bigg) \bigg),
\end{align}
where $\sum_{\abs 1}$ simply means summing over the terms with positive and negative longitudinal directions.
We now have two sets of equations: one for longitudinal components, and two for transverse components.

The equations of motion can be written more compactly by introducing the symbol
\bb
K_{x,ij}[U, C] = - \frac{1}{2} \frac{1}{\lb a^i a^j \rb^2} \lb U_{x+i,j} C^\dg_{x,ij} - C^\dg_{x-j,ij} U_{x-j,j}\rb,
\ee
where $C$ can be exchanged for corresponding expressions with $M$ or $W$ and $U$ can be exchanged for its temporally averaged version $\avg{U}$. The longitudinal component of the EOM then reads
\begin{align}
\frac{1}{\lb a^{0}a^{1} \rb^{2}} P^{a} \lb U_{x,10}+U_{x,1-0} \rb =  \frac{1}{2} \sum_{\abs i} P^a \lb U_{x,1} \lb K_{x,1i}[U, W] + \avg{K}_{x,1i}[U, C] \rb \rb,
\end{align}
and the transverse components are given by
\begin{align}
\frac{1}{\lb a^{0}a^{i} \rb^{2}} P^{a} \lb U_{x,i0}+U_{x,i-0} \rb = P^a \bigg( U_{x,i} \bigg( &\sum_{\abs j} K_{x,ij}[U, M] \nonumber \\
&+ \frac{1}{2} \sum_{\abs 1} \lb K_{x,i1}[U, W] + K_{x,i1}[\avg{U}, C] \rb \bigg) \bigg).
\end{align}
These equations can be solved numerically using damped fixed point iteration completely analogously to section \ref{implicit_scheme}. First, one obtains an initial guess $U^{(0)}_{x+0,i}$ for ``future" link variables from a single leapfrog evolution step using eqs.\ \eqref{eq:leapfrog_eom} and \eqref{eq:link_evolve}. Then iterate from $n=1$ until convergence:
\begin{enumerate} 
\item Compute the next iteration using damped fixed point iteration: in eqs.\ \eqref{eq:semi_eom_1} and \eqref{eq:semi_eom_i} replace $P^a \lb U_{x,10}\rb \rightarrow \mathcal{U}^a_1$ and $P^a \lb U_{x,i0} \rb \rightarrow \mathcal{U}^a_i$, solve for the unknown $\mathcal U$'s and update the temporal plaquettes using
\bb
P^a \lb U^{(n)}_{x,10} \rb = \alpha P^a \lb  U^{(n-1)}_{x,10} \rb + \lb 1 - \alpha \rb \mathcal U^{a}_1
\ee
and analogously for $U^{(n)}_{x,i0}$ and $\mathcal U^{a}_i$. $\alpha$ is the damping coefficient.
\item For SU(2) we can reconstruct the full temporal plaquette from its components  $P^a \lb U \rb$ with the identity
\bb
U = \sqrt{1-\sum_a  P^a \lb U \rb ^2} \one + \frac{i}{2} \sum_a \sigma^a P^a \lb U \rb,
\ee
for $U = U^{(n)}_{x,10}$ and $U = U^{(n)}_{x,i0}$. 
\item Using $U^{(n)}_{x,10}$ and $U^{(n)}_{x,i0}$, compute the spatial links $U^{(n)}_{x+0,1}$ and $U^{(n)}_{x+0,i}$ via
\bb
U^{(n)}_{x+0,i} = U^{(n)}_{x,0i} U_{x,i}.
\ee
\item Repeat with $n\rightarrow n+1$.
\end{enumerate}
As with the implicit scheme, our approach to solving the equations in the semi-implicit scheme is an iterative one. The Gauss constraint \eqref{eq:semi_gauss} is only approximately satisfied, depending on the degree of convergence\footnote{This is also true for the Abelian semi-implicit scheme. If the equations of motion are solved only approximately using an iterative method, then the conservation of the Gauss constraint is also only approximate depending on the degree of convergence.}.

\subsection{Coupling to external color currents} \label{external_charges}

Up until now we have only considered pure Yang-Mills fields. In the continuum we can include external color currents by adding a $J \cdot A$ term to the action.
\bb
S[A] = S_{YM} + S_J = -\frac{1}{2} \intop_x \sum_{\mu, \nu} \tr \lb F_{\mu \nu}(x) F^{\mu \nu}(x) \rb - 2 \intop_x \sum_{\nu} \tr \lb J^\nu(x) A_{\nu}(x) \rb. 
\ee
The equations of motion then read
\bb
\sum_\mu D_{\mu} F^{\mu\nu}(x) = J^\nu(x),
\ee
and due to gauge-covariant conservation of charge we have
\bb
\sum_\mu D_{\mu} J^{\mu}(x)=0,
\ee
which is the non-Abelian continuity equation. On the lattice we can simply add a discrete $J\cdot A$ term to the action as well:
\bb
S_J = \frac{V}{g^2} \sum_{x,b} \lb - \frac{g}{a^0} \rho^b_x A^b_{x,0} + \sum_{i=1}^3 \frac{g}{a^i} j^b_{x,i} A^b_{x,i} \rb,
\ee
where $A^a_{x,\mu}$ includes a factor of $g a^\mu$ (``lattice units"). We also made the split into 3+1 dimensions explicit using $J^a_0(x) \simeq \rho^a_x$ and $J^a_i(x) \simeq j^a_{x,i}$. The variation of $S_J$ simply reads
\bb
\dd S_J = \frac{V}{g^2} \sum_{x,b} \lb - \frac{g}{a^0} \rho^b_x \dd A^b_{x,0} + \sum_{i=1}^3 \frac{g}{a^i} j^b_{x,i} \dd A^b_{x,i} \rb,
\ee
which gives the appropriate contributions to the Gauss constraint and the EOM.
In the leapfrog scheme the Gauss constraint now reads
\bb \label{eq:leapfrog_gauss_current}
\sum_i \frac{1}{\lb a^0 a^i \rb^2} P^a \lb U_{x,0i} + U_{x,0-i}\rb = \frac{g}{a^0} \rho^a_x .
\ee
and the EOM read
\bb\label{eq:leapfrog_eom1_current}
 \frac{1}{\lb a^0 a^i \rb^2} P^a \lb U_{x,i0} + U_{x,i-0}\rb =
\sum_j \frac{1}{\lb a^i a^j \rb^2} P^a \lb U_{x,ij} + U_{x,i-j}\rb - \frac{g}{a^i} j^a_{x,i}.
\ee
The constraint taken together with the EOM imply the local conservation of charge (see appendix \ref{app_gauss})
\bb \label{eq:continuity_discrete}
\frac{\rho_x - \rho_{x-0}}{a^0} = \sum_i \frac{j_{x,i} - U^\dg_{x-i,i} j_{x-i,i} U_{x-i,i}}{a^i},
\ee
which is the discrete version of the continuity equation. For the implicit and semi-implicit schemes the procedure is the same: including the $S_J$ term simply leads to the appearance of $\rho$ on the RHS of the Gauss constraint (see eqs.\ \eqref{eq:implicit_gauss} and \eqref{eq:semi_gauss}) and $j_{x,i}$ on the RHS of the EOM (see eqs.\ \eqref{eq:implicit_eom} and \eqref{eq:semi_eom_1}, \eqref{eq:semi_eom_i}). Due to conservation of the Gauss constraint without external charges, the continuity equation for the implicit and semi-implicit scheme is simply eq.\ \eqref{eq:continuity_discrete} as well. This implies that our treatment of the external currents in terms of parallel transport as detailed in our previous publication \citep{Gelfand:2016prm} does not require any modifications when using the newly derived schemes.

\section{Numerical tests}\label{tests}

In this last section we test the semi-implicit scheme on the propagation of a single nucleus in the CGC framework. For an observer at rest in the laboratory frame, the nucleus moves at the speed of light and consequently exhibits large time dilation. As the nucleus propagates the interactions inside appear to be frozen and the field configuration is essentially static.
On the lattice we would like to reproduce this behavior as well, but depending on the lattice resolution we run into the numerical Cherenkov instability, which leads to an artificial increase of the total field energy of the system.

\begin{figure}[tbp]
	\centering
	\includegraphics{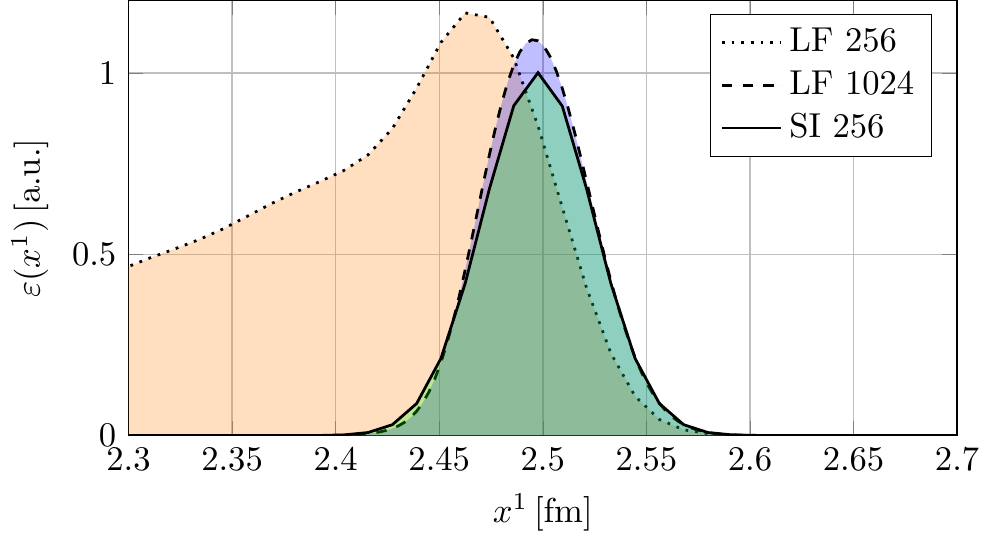}
	\caption{
		The energy density of a right-moving nucleus averaged over the transverse plane as a function of the longitudinal coordinate $x^1$ after $t=2\,\text{fm}/c$. The results were obtained from the same simulations as in figure \ref{fig:increase}. We compare the performance of the leapfrog (LF) scheme (with $N_L=256$ and $N_L=1024$) to the semi-implicit (SI) scheme with $N_L=256$. In the most extreme example (LF 256) the nucleus becomes completely unstable due to the numerical Cherenkov instability. By eliminating numerical dispersion (SI 256) the nucleus retains its original shape almost exactly.
		\label{fig:profiles}}
\end{figure}

As previously stated the root cause of the instability is numerical dispersion: in the CGC framework a nucleus consists of both propagating field modes and a longitudinal current generating the field around it. It is essentially a non-Abelian generalization of the field of a highly relativistic electric charge.
In our simulation the color current is modeled as an ensemble of colored point-like particles moving at the speed of light along the beam axis. The current is unaffected by dispersion, i.e.\ it retains its shape perfectly as it propagates. The field modes suffer from numerical dispersion, which over time leads to a deformation of the original longitudinal profile. The mismatch between the color current and the field leads to creation of spurious field modes, which interact with the color current non-linearly through parallel transport (color rotation) of the current. This increases the mismatch further and more spurious fields are created. As the simulation progresses this eventually leads to a large artificial increase of total field energy. The effects of the instability can be quite dramatic as seen in figure \ref{fig:profiles}.
The main difference to the numerical Cherenkov instability in Abelian PIC simulations is that in electromagnetic simulations the spurious field modes interact with the particles through the Lorentz force \citep{GODFREY1974504}. In our simulations we do not consider any acceleration of the particles (i.e.\ their trajectories are fixed), but interaction is still possible due to non-Abelian charge conservation \eqref{eq:continuity_discrete}, which requires rotating the color charge of the color current. Therefore our type of numerical Cherenkov instability is due to non-Abelian effects.

We now demonstrate that the instability is cured (or at least highly suppressed for all practical purposes) by using the semi-implicit scheme. We test the schemes ability to improve energy conservation the following way: we place a single gold nucleus described by the McLerran-Venugopalan model in a simulation box of volume $V = \lb 3\,\text{fm}\rb \times \lb 6\,\text{fm}\rb^2$ and set the longitudinal extent of the nucleus to roughly correspond to a boosted nucleus with Lorentz factor $\gamma = 100$. These parameters correspond to a similar setting as in our previous work \citep{Ipp:2017lho}. We refer to \citep{Gelfand:2016yho} for a detailed description of the initial conditions. After setting up the initial condition we let the nucleus freely propagate along the longitudinal axis. As the simulation runs we record the total field energy
\bb
E(t) = \frac{1}{2} \int_V d^3 x \sum_{i,a} \lb  E^a_i\lb t, \vec{x}\rb^2 \ +  B^a_i\lb t, \vec{x}\rb^2 \rb,
\ee
where $E^a_i(x)$ and $B^a_i(x)$ are the color-electric and -magnetic fields at each time step. On the lattice the electric and magnetic fields are approximated using plaquettes:
\begin{align}
E^a_{i}(x) &\simeq \frac{1}{g a^0 a^i} P^a \lb U_{x,0i} \rb, \\
B^a_{i}(x) &\simeq -\sum_{j, k} \varepsilon_{ijk} \frac{1}{2 g a^j a^k} P^a \lb U_{x,jk} \rb.
\end{align}
We compute the relative change $\lb E(t) - E(0) \rb / E(0)$, which we plot as a function of time $t$. In the continuum we would have $E(t)=E(0)$, but due to numerical artifacts and the Cherenkov instability this is not the case in our simulations.

\begin{figure}[tbp]
	\centering
	\includegraphics{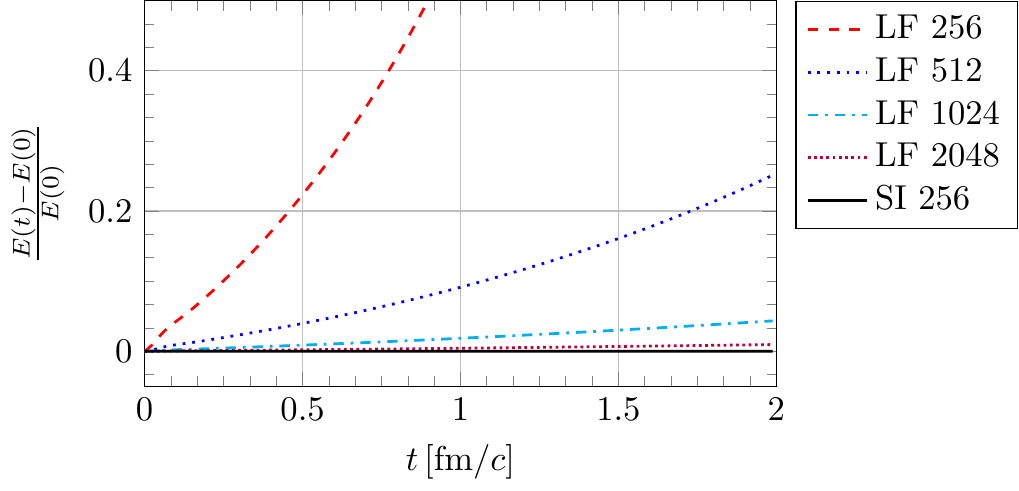}
	\caption{
		The relative increase of the total field energy $E(t)$ as a function of time $t$ for the propagation of a single nucleus in a box of volume $V = \lb 3\,\text{fm}\rb \times \lb 6\,\text{fm}\rb^2$ with a longitudinal length of $3\,\text{fm}$ and a transverse area of $\lb 6 \, \text{fm}\rb^2$. The lattice size is $N_L \times N_T^2$ with the transverse lattice fixed at $N_T = 256$. Starting with the same initial condition, we evolve forward in time using the leapfrog (LF) and the semi-implicit (SI) method. In the case of the leapfrog scheme we vary the resolution along the beam axis using the number of longitudinal cells $N_L$ of the lattice. For $N_L=256$ the numerical Cherenkov instability leads to catastrophic failure, increasing the energy to many times its original value. The effect is suppressed when increasing the longitudinal resolution, but the instability is still present. In the case of the semi-implicit scheme it is possible to set $N_L=256$ and still obtain (approximate) energy conservation. After $t=2\,\text{fm}/c$ the energy increase for the semi-implicit scheme is roughly $0.02\%$, compared to $1\%$ for the leapfrog with $N_L=2048$. For the simulation using the semi-implicit scheme we used $N_i = 10$ iterations and a damping coefficient of $\alpha=0.45$. The time step is set to the longitudinal lattice spacing. In the case of the leapfrog simulation we used $a^0 = a^1 / 4$. 
		\label{fig:increase}}
\end{figure}

The numerical results are shown in figure \ref{fig:increase}. We see that the leapfrog scheme leads to an exponential increase of the total energy over time, which can be suppressed using finer lattices. On the other hand, the semi-implicit scheme leads to better energy conservation even on a rather coarse lattice. Therefore, the resolution that is usually required to obtain accurate, stable results is lowered by using the semi-implicit scheme. However, using the new scheme might not always be economical: finer lattices suppress the instability as well and since the leapfrog scheme is computationally cheaper than the semi-implicit scheme, the leapfrog can be favorable in practice. On our test system (a single $256$ gb node on the VSC 3 cluster) the simulation using the semi-implicit scheme (SI 256) takes $\sim 4$ hours to finish, while the same simulation using the leapfrog with $N_L=1024$ (LF 1024) takes roughly $\sim 2.5$ hours and with $N_L=2048$ (LF 2048) $\sim 10$ hours. Even though energy conservation is not as good as SI 256, the longitudinal resolution is much better in comparison enabling us to extract observables with higher accuracy. It should be noted however that our implementation of the leapfrog scheme is already highly optimized, while the implementation of the semi-implicit scheme is very basic and should be considered as a proof of concept. Further optimizations and simplifications of the semi-implicit scheme might make it the better choice in most cases. 

\begin{figure}[tbp]
\centering
\includegraphics{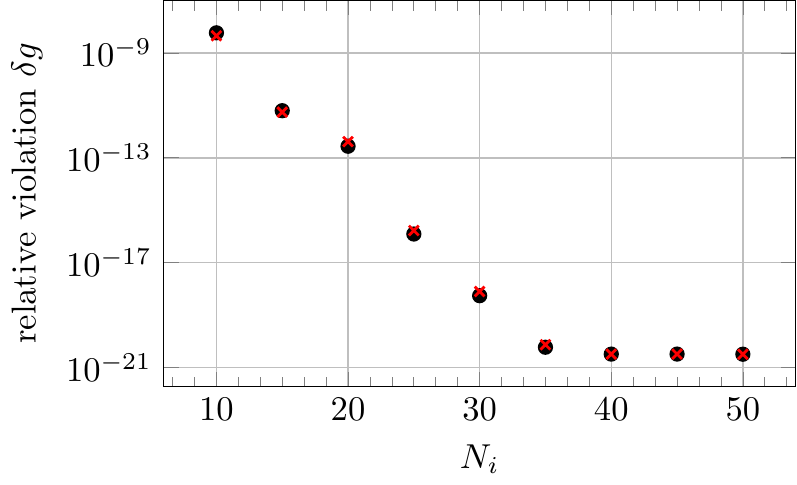}
\caption{
The relative Gauss constraint violation for the semi-implicit scheme as a function of the number of iterations $N_i$ of the damped fixed point iteration. For this plot we used a simulation box of volume $V = \lb 1.5\,\text{fm}\rb \times \lb 6\,\text{fm} \rb^2$ on a lattice with $N_T=256$ points in the transverse directions and $N_L=128$ points in the longitudinal direction. Otherwise, we use the same initial conditions as in figures \ref{fig:increase} and \ref{fig:profiles}. At the beginning of the simulation at $t=0\,\text{fm}/c$ the constraint is conserved up to machine precision by construction. We then let the nucleus propagate until $t=0.5\,\text{fm}/c$ and compute the violation of the Gauss constraint (black dots). We also compare to the constraint violation after only a single evolution step (red crosses), which does not differ much from the violation after a larger number of time steps. It is evident that the violation systematically converges towards zero (up to machine precision) as we increase the number of iterations $N_i$.
\label{fig:gauss}}
\end{figure}

As a second test we look at the violation of the Gauss constraint. The leapfrog scheme \eqref{eq:leapfrog_eom} conserves its associated Gauss constraint \eqref{eq:leapfrog_gauss} identically, even for finite time-steps $a^0$. In numerical simulations this conservation is not exact due to floating point number errors, but the violation is zero up to machine precision.
On the other hand, the semi-implicit scheme has to be solved iteratively and therefore the results depend on the number of iterations $N_i$ used in the fixed point iteration method. 
We define the relative violation of the Gauss constraint as the ratio of the absolute (squared) Gauss constraint violation to the total (squared) charge on the lattice. In the case of the leapfrog scheme this reads
\bb \label{eq:rel_gauss}
\dd g(t) = \frac{\sum_{x',a} \lb \sum_i \frac{1}{\lb a^0 a^i \rb^2} P^a \lb U_{x',0i} + U_{x',0-i}\rb - \frac{g}{a^0} \rho^a_{x'}\rb^2}{\sum_{x',a} \lb \frac{g}{a^0} \rho^a_{x'} \rb^2},
\ee
where the sum $\sum_{x'}$ runs over the spatial lattice of a single time slice at $t$. The numerator depends on the Gauss constraint of the scheme and has to be adjusted according to the implicit and semi-implicit method (either eq.\ \eqref{eq:implicit_gauss} or \eqref{eq:semi_gauss} including the charge density on the RHS as discussed in section \ref{external_charges}). 
In figure \ref{fig:gauss} we show how the Gauss constraint violation converges systematically towards zero as we increase the number of iterations. Therefore, even though we can not use an arbitrarily high number of iterations due to limited computational resources, the semi-implicit scheme conserves the Gauss constraint in principle. The same holds for the purely implicit scheme. In practice it is not necessary to satisfy the constraint up to high precision, as observables such as the energy density seem to converge much faster up to satisfying accuracy.

\section{Conclusions and Outlook}

In this paper we derived new numerical schemes for real-time lattice gauge theory. We started our discussion based on two simpler models, namely the two-dimensional wave equation and Abelian gauge fields on the lattice. It turns out that using a discrete variational principle to derive numerical schemes for equations of motion is a very powerful tool: the use of time-averaged expressions in the discrete action yields implicit and semi-implicit schemes depending on how exactly the time-averaging is performed and what terms are replaced by their averages.
We extended this concept to real-time lattice gauge theory, allowing us to make modifications to the standard Wilson gauge action that yield new numerical schemes, which have the same accuracy as the leapfrog scheme and, most importantly, are gauge-covariant and conserve the Gauss constraint. Finally, we demonstrated a peculiar property of the semi-implicit scheme: it allows for dispersion-free propagation along one direction on the lattice, thus curing a numerical instability that has plagued our simulations of three-dimensional heavy-ion collisions.

Although all numerical tests in this work have been performed for the propagation of a single nucleus, we expect that the new semi-implicit scheme will improve simulations of nucleus-nucleus collisions in multiple ways. Primarily, using the new scheme we can be sure that the color fields of incoming nuclei have not been altered up until the collision event and all changes to the fields afterwards are solely due to interaction between the colliding nuclei during the collision event itself. This also helps to improve numerical accuracy in the forward and backward rapidity region: at later simulation times, when the now outgoing nuclei are well separated, all field modes (i.e.\ ``gluons") with momenta at high rapidity must have been created in the collision and contributions from artificial modes emitted by the nucleus due to numerical Cherenkov radiation are strongly suppressed. Furthermore, from the dispersion relations \eqref{eq:semi_dispersion1} and \eqref{eq:semi_dispersion2} we can infer that these gluons with almost purely longitudinal momentum $k_L$ (and small transverse momentum $k_T$) exhibit a phase velocity approximately the speed of light (up to an error term quadratic in  $a^0 k_T$). This means any interactions between high rapidity gluons produced in the collision and the finite-thickness color fields of the nuclei directly after the collision event can be considered physical and are not tainted by numerical dispersion as in our previous simulations. Consequently, it should be possible to extract space-time rapidity profiles of the local rest frame energy density of the Glasma as in \cite{Ipp:2017lho} valid for larger ranges of rapidity as previously considered.  

In conclusion, we hope that this new treatment of solving the Yang-Mills equations on the lattice allows us to perform better, more accurate simulations using more complex models of nuclei.

\begin{acknowledgments}

The authors thank A.\ Dragomir, C.\ Ecker, T.\ Lappi, J.\ Peuron and A.\ Polaczek
for helpful discussions and comments. This work has been supported by the Austrian Science Fund FWF, Project No.\ P26582-N27 and Doctoral program No.\ W1252-N27.\ The computational results have been achieved using the Vienna Scientific Cluster.

\end{acknowledgments}

\appendix

\section{Stability analysis of the semi-implicit scheme for Abelian gauge fields}\label{app_abelian_semi}

In this section of the appendix we prove stability for the semi-implicit scheme for Abelian fields derived in section \ref{sec:abelian_semi_implicit}. Using temporal gauge, $A_{x,0}=0$, the equations of motion \eqref{eq:abelian_eom_d3_1} and \eqref{eq:abelian_eom_d3_2} read
\begin{align}
-\p_{0}^{2}A_{x,1}	&=	\frac{1}{2}\sum_{i}\p_{i}^{B}\left(W_{x,1i}+M_{x,1i}\right), \\
-\p_{0}^{2}A_{x,i}	&=	\sum_{j\neq i}\p_{j}^{B}M_{x,ij}+\frac{1}{2}\p_{1}^{B}\left(W_{x,i1}+F_{x,i1}\right),
\end{align}
where $W_{x,1i}$ reduces to
\bb
W_{x,1i}=\p_{1}^{F}A_{x,i}-\p_{i}^{F}\bar{A}_{x,1}.
\ee
The Gauss constraint \eqref{eq:abelian_constraint_d3} reads
\bb
\sum_{i=1}^{d}\p_{i}^{B}\p_{0}^{F}A_{x,i}+\left(\frac{a^{0}}{2}\right)^{2}\sum_{i}\p_{1}^{B}\p_{i}^{B}\p_{0}^{F}\left(\p_{1}^{F}A_{x,i}-\p_{i}^{F}A_{x,1}\right)=0.
\ee
Splitting the equations of motion into Laplacian terms and mixed derivative terms we find
\begin{align}
-\p_{0}^{2}A_{x,1}	&=	-\sum_{i}\p_{i}^{2}\avg{A}_{x,1}
+\frac{1}{2}\sum_{i}\p_{i}^{B}\p_{1}^{F}\left(A_{x,i}+\avg{A}_{x,i}\right), \\
-\p_{0}^{2}A_{x,i}	&=	-\sum_{j\neq i}\p_{j}^{2}\avg{A}_{x,i}-\p_{1}^{2}A_{x,i}
+\sum_{j\ne i}\p_{j}^{B}\p_{i}^{F}\avg{A}_{x,j}+\frac{1}{2}\p_{1}^{B}\p_{i}^{F}\left(\avg{A}_{x,1}+A_{x,1}\right).
\end{align}
Using a plane wave ansatz
\bb
A_{x,i} = A_i e^{i \lb \omega x^0 - \sum_i k^i x^i \rb},
\ee
we will use the Gauss constraint to first reduce the number of degrees of freedom and then compute the dispersion relation $\omega(k)$. Inserting the ansatz into the Gauss constraint we find
\bb
\left(1+\sum_{i}\chi_{i}^{2}\right)\chi_{1}^{B}A_{1}+\left(1-\chi_{1}^{2}\right)\sum_{i}\chi_{i}^{B}A_{i}=0,
\ee
where we use the dimensionless lattice momenta \eqref{eq:dimless_lattice_momentum}. The constraint equation can alternatively be written as
\bb
\chi_{1}^{B}A_{1}=-\beta\sum_{i}\chi_{i}^{B}A_{i},
\ee
where $\beta$ is a momentum-dependent factor given by
\bb
\beta = \frac{1-\chi_{1}^{2}}{1+\sum_{i}\chi_{i}^{2}}.
\ee
The temporal average $\avg{A}_{x,i}$ reduces to a multiplication with a frequency dependent factor:
\bb
\avg{A}_{x,i} = \cos \lb \omega a^0 \rb A_{x,i} = c A_{x,i},
\ee
where we used the shorthand $c = \cos \lb \omega a^0 \rb$. Inserting the plane wave ansatz into the EOM yields
\begin{align}
\chi_{0}^{2}A_{1} &= c \sum_{i} \chi_{i}^{2} A_{1} - \frac{1}{2} \left( 1+c \right) \sum_{i} \chi_{i}^{B} \chi_{1}^{F} A_{i}, \\
\chi_{0}^{2}A_{i} &= c \sum_{j\neq i} \chi_{j}^{2} A_{i} + \chi_{1}^{2} A_{i} - c \sum_{j\ne i} \chi_{j}^{B} \chi_{i}^{F} A_{j} - \frac{1}{2}\left(1+c\right)\chi_{1}^{B}\chi_{i}^{F}A_{1}.
\end{align}
Note that both $\chi_0$ and $c$ depend on $\omega$. After making use of the Gauss constraint the longitudinal EOM reads
\bb
\chi_{0}^{2}A_{1}=c\left(\chi_{2}^{2}+\chi_{3}^{2}\right)A_{1}+\frac{1}{2}\left(1+c\right)\beta^{-1}\chi_{1}^{2}A_{1},
\ee
and the two transverse equations read
\begin{align}
\chi_{0}^{2}A_{2}	&=	c\chi_{3}^{2}A_{2}+\chi_{1}^{2}A_{2}-c\chi_{3}^{B}\chi_{2}^{F}A_{3}-\frac{1}{2}\left(1+c\right)\chi_{1}^{B}\chi_{2}^{F}A_{1} \nonumber \\
&=	c \chi_{3}^{2}A_{2}+\chi_{1}^{2}A_{2}+\frac{1}{2}\left(1+c\right)\beta\chi_{2}^{2}A_{2}+\left(\frac{1}{2}\left(1+c\right)\beta-c\right)\chi_{2}^{F}\chi_{3}^{B}A_{3}, \\
\chi_{0}^{2}A_{3}	&=	c\chi_{2}^{2}A_{3}+\chi_{1}^{2}A_{3}-c\chi_{2}^{B}\chi_{3}^{F}A_{2}-\frac{1}{2}\left(1+c\right)\chi_{1}^{B}\chi_{3}^{F}A_{1} \nonumber \\
&=	c\chi_{2}^{2}A_{3}+\chi_{1}^{2}A_{3}+\frac{1}{2}\left(1+c\right)\beta\chi_{3}^{2}A_{3}+\left(\frac{1}{2}\left(1+c\right)\beta-c\right)\chi_{2}^{B}\chi_{3}^{F}A_{2}.
\end{align}
This system of equations can be written in matrix notation as an eigenvalue problem
\bb
M \vec{A} = \chi^2_0 \vec{A},
\ee
where the coefficient matrix $M$ is given by
\bb
M = \left(\begin{array}{ccc}
	\frac{1}{2}\left(1+c\right)\beta^{-1}\chi_{1}^{2}+c\left(\chi_{2}^{2}+\chi_{3}^{2}\right) & 0 & 0\\
	0 & \chi_{1}^{2}+\frac{1}{2}\left(1+c\right)\beta\chi_{2}^{2}+c\chi_{3}^{2} & \left(\frac{1}{2}\left(1+c\right)\beta-c\right)\chi_{2}^{F}\chi_{3}^{B}\\
	0 & \left(\frac{1}{2}\left(1+c\right)\beta-c\right)\chi_{2}^{B}\chi_{3}^{F} & \chi_{1}^{2}+c\chi_{2}^{2}+\frac{1}{2}\left(1+c\right)\beta\chi_{3}^{2}
\end{array}\right),
\ee
and the vector $\vec{A}$ is simply
\bb
\vec{A} =  \lb \begin{array}{c}
	A_{1}\\
	A_{2}\\
	A_{3}
\end{array} \rb.
\ee
The eigenvectors of $M$ are
\bb
\left\{ \vec{A}_{L},\vec{A}_{T,1},\vec{A}_{T,2}\right\} =
\left\{
\left(
\begin{array}{c}
	1\\
	0\\
	0
\end{array}
\right),
\left(
\begin{array}{c}
	0\\
	-\chi_{3}^{B}\\
	\chi_{2}^{B}
\end{array}
\right),
\left(
\begin{array}{c}
	0\\
	\chi_{2}^{F}\\
	\chi_{3}^{F}
\end{array}
\right)\right\}, 
\ee
where we find two transverse, momentum dependent eigenvectors $\vec{A}_{T,1}$, $\vec{A}_{T,2}$ and the longitudinal unit vector $\vec{A}_L$. The two transverse vectors $\vec{A}_{T,1}$ and $\vec{A}_{T,2}$ can be interpreted as transverse polarization modes and are orthogonal in the sense of $\left(\vec{A}_{T,1}\right)\cdot\left(\vec{A}_{T,2}\right)^{\dg}=0$. The three eigenvectors yield three different equations for the eigenvalue problem, namely
\begin{align}
M \vec{A}_L &= \lambda_L(c, \chi_i) \vec{A}_L = \chi^2_0 \vec{A}_L, \\
M \vec{A}_{T,l} &= \lambda_{T,l}(c, \chi_i) \vec{A}_{T,l} = \chi^2_0 \vec{A}_{T,l}, \quad l \in {1, 2},
\end{align}
where $\lambda_L(c, \chi_i)$ and $\lambda_{T,k}(c, \chi_i)$ are expressions which depend on the momenta $\chi_i$ and the frequency $\omega$ via $c=\cos{\lb \omega a^0 \rb}$.
Solving the first equation $\lambda_L(c, \chi_i) = \chi^2_0$ for $\omega$ yields the dispersion relation of the longitudinal component:
\bb \label{eq_app_semi_freq1}
\omega_{L}a^{0}=\arccos\left(\frac{1-\chi_{1}^{2}\left(2+\chi_{2}^{2}+\chi_{3}^{2}\right)}{1+\chi_{2}^{2}\left(2-\chi_{1}^{2}\right)+\chi_{3}^{2}\left(2-\chi_{1}^{2}\right)}\right).
\ee
Solving the two transverse equations $\lambda_{T,l}(c, \chi_i) = \chi^2_0$ yields
\bb \label{eq_app_semi_freq2}
\omega_{T,1}a^{0}=\arccos\left(\frac{1-2\chi_{1}^{2}}{1+2\chi_{2}^{2}+2\chi_{3}^{2}}\right),
\ee
and
\bb
\omega_{T,2}a^{0}=\arccos\left(\frac{1-\chi_{1}^{2}\left(2+\chi_{2}^{2}+\chi_{3}^{2}\right)}{1+\chi_{2}^{2}\left(2-\chi_{1}^{2}\right)+\chi_{3}^{2}\left(2-\chi_{1}^{2}\right)}\right)=\omega_{L}a^{0}.
\ee
It turns out that the expressions for $\omega$ associated with the different eigenvectors are not the same, although $\vec{A}_L$ and $\vec{A}_{T,2}$ share the same dispersion relation. We interpret this as numerical (or artificial) birefringence. Given a momentum $k$, the amplitude of an arbitrary wave has to be split into two components: a part which is projected into the plane spanned by $\vec{A}_L$ and $\vec{A}_{T,2}$ which oscillates with $\omega_L=\omega_{T,2}$ and a part parallel to $\vec{A}_{T,1}$ which oscillates with frequency $\omega_{T,1}$.

We require the propagation of a wave to be stable, i.e.\ we require the frequencies $\omega$ to be real-valued. This is guaranteed if the arguments of the $\arccos$ expressions in eqs.\ \eqref{eq_app_semi_freq1} and \eqref{eq_app_semi_freq2} are restricted to $[-1,1]$. Both dispersion relations remain stable if the CFL condition
\bb
\chi^2_1 \leq 1,
\ee
holds. Using 
\bb
\chi^2_1 = \lb \frac{a^0}{a^1} \rb \sin^2 \lb \frac{k^1 a^1}{2} \rb,
\ee
and requiring stability for all values of $k^1$ yields the constraint
\bb
a^0 \leq a^1.
\ee
This concludes the proof that the semi-implicit scheme, even though exhibiting peculiar wave propagation phenomena, is stable.

\section{Variation of gauge links}\label{app_var}

We introduce the infinitesimal variation of a gauge link variable
\bb
\dd U_{x,\mu} = i \dd A_{x,\mu} U_{x,\mu},
\ee
where the variation of the gauge field $\dd A_{x,\mu}$ is traceless and hermitian. In the continuum limit $\dd A_{x,\mu}$ becomes the infinitesimal variation of the gauge field $A_{\mu}(x)$.
The infinitesimal variation $\dd U_{x,\mu}$ preserves the unitary of gauge links. Let $U'_{x,\mu} = U_{x,\mu} + \dd U_{x,\mu}$ then we find
\begin{align}
U'_{x,\mu} U'^\dg_{x,\mu} &= \lb U_{x,\mu} + \dd U_{x,\mu} \rb \lb U^\dg_{x,\mu} + \dd U^\dg_{x,\mu} \rb \nonumber \\
& \simeq \one + \dd U_{x,\mu} U^\dg_{x,\mu} + U_{x,\mu} \dd U^\dg_{x,\mu} +\mathcal O \lb \dd A^2 \rb \nonumber \\
& \simeq \one + i \dd A_{x,\mu} - i \dd A_{x\,\mu } +\mathcal O \lb \dd A^2 \rb \nonumber \\
& \simeq \one + \mathcal O \lb \dd A^2 \rb.
\end{align}
The determinant is also unaffected for infinitesimal variations.
\begin{align}
\det U'_{x,\mu} &= \det \lb U_{x,\mu} + \dd U_{x,\mu} \rb  \nonumber \\
& \simeq 1 + \tr \lb \adj \lb U_{x,\mu} \rb \dd U_{x,\mu} \rb + \mathcal O \lb \dd A^2 \rb \nonumber \\
& \simeq 1 + \tr \lb U^\dg_{x,\mu} \dd U_{x,\mu} \rb + \mathcal O \lb \dd A^2 \rb \nonumber \\
& \simeq 1 + \mathcal O \lb \dd A^2 \rb.
\end{align}
The variation $\dd U_{x,\mu}$ therefore preserves the constraints and allows us to vary the action without Lagrange multipliers, which dramatically simplifies the derivation of equations of motion.

\section{Variation of the leapfrog action}\label{app_leapfrog}
In this section of the appendix we give a derivation of the discrete equations of motion obtained from the standard Wilson action \eqref{eq:wilson_action} using the constraint preserving variation of link variables of the previous section. To make the calculation more organized, we first split the action into two parts: a part containing temporal plaquettes $S_E[U]$ (``E" for electric) and a part containing spatial plaquettes $S_B[U]$ (``B" for magnetic). We write
\bb
S[U] = S_E[U] - S_B[U],
\ee
where
\begin{align}
S_E[U] &= \frac{V}{g^2} \sum_{x,i} \frac{1}{\lb a^0 a^i \rb^2} \tr \lb C_{x,0i} C^\dg_{x,0i} \rb,\\
S_B[U] &= \frac{V}{g^2} \sum_{x,i,j} \frac{1}{2} \frac{1}{\lb a^i a^j \rb^2} \tr \lb C_{x,ij} C^\dg_{x,ij} \rb.
\end{align}
\subsection{Gauss constraint} \label{app_leapfrog_gauss}
For the Gauss constraint in the leapfrog scheme we only have to consider the variation $\dd_t S_E[U]$ as $S_B[U]$ does not contain any temporal links.
In the following sections we make use of the ``$\sim$" symbol, denoting equality under the sum over lattice sites $x$ and under the trace.
We then have
\begin{align}
\dd_t \lb C_{x,0i} C^\dg_{x,0i} \rb &= \lb \dd U_{x,0} U_{x+0,i} - U_{x,i} \dd U_{x+i,0} \rb C^\dg_{x,0i} + \hc \nonumber \\
&\sim \dd U_{x,0} \lb U_{x+0,i} C^\dg_{x,0i} - C^\dg_{x-i,0i} U_{x-i,i}\rb + \hc \nonumber \\
&=i \dd A_{x,0} \lb U_{x,0} \lb U_{x+0,i} C^\dg_{x,0i} - C^\dg_{x-i,0i} U_{x-i,i} \rb \rb + \hc
\end{align}
To go from the first to the second line, we applied a shift $x \rightarrow x-i$ in the right term of the first line and made use of the cyclicity of the trace. In the third line we simply used the definition of the variation of gauge links.
The variation of the action therefore reads
\bb
\dd_t S_E[U] = -\frac{V}{g^{2}}\sum_{x,i,a}\frac{1}{\left(a^{0}a^{i}\right)^{2}}P^{a}\left(U_{x,0}\left(U_{x+0,i}C_{x,0i}^{\dg}-C_{x-i,0i}^{\dg}U_{x-i,i}\right)\right)\delta A_{x,0}^{a},
\ee
where we used
\bb
P^a\lb U \rb \equiv 2 \Im \tr \lb t^a U \rb = -i \tr \lb t^a \lb U - U^\dg \rb \rb.
\ee
Replacing the $C_{x,ij}$ terms with link variables we find
\bb
U_{x,0}\left(U_{x+0,i}C_{x,0i}^{\dg}-C_{x-i,0i}^{\dg}U_{x-i,i}\right) = 2-U_{x,0i}-U_{x,0-i},
\ee
and subsequently
\bb
\dd_t S_E[U] = \frac{V}{g^{2}}\sum_{x,i,a} \frac{1}{\left(a^{0}a^{i}\right)^{2}}P^{a}\left(U_{x,0i}+U_{x,0-i}\right)\delta A_{x,0}^{a}.
\ee
We require that the variation vanishes, i.e.\ $\dd S[U] = 0$. Since all gauge links can be varied independently we find
\bb
\sum_{i}\frac{1}{\left(a^{0}a^{i}\right)^{2}}P^{a}\left(U_{x,0i}+U_{x,0-i}\right)=0.
\ee
\subsection{Equations of motion} \label{app_leapfrog_eom}

First we consider the variation of $S_E[U]$ w.r.t\ spatial links. We find a result that is similar to the expression for the Gauss constraint
\bb
\delta_{s}\left(C_{x,0i}C_{x,0i}^{\dg}\right) \sim i\delta A_{x,i}\left(U_{x,i}\left(C_{x-0,0i}^{\dg} U_{x-0,0}-U_{x+i,0} C_{x,0i}^{\dg}\right)\right)+ \hc
\ee
and
\begin{align}
\dd_s S_E[U] &= -\frac{V}{g^{2}}\sum_{x,i,a}\frac{1}{\left(a^{0}a^{i}\right)^{2}} \dd A_{x,i}^{a}  P^{a}\left(U_{x,i}\left(C_{x-0,0i}^{\dg}U_{x-0,0}-U_{x+i,0}C_{x,0i}^{\dg}\right)\right)\nonumber \\
&= \frac{V}{g^{2}}\sum_{x,i,a}\frac{1}{\left(a^{0}a^{i}\right)^{2}} \dd A_{x,i}^{a}  P^{a}\left(U_{x,i0}+U_{x,i-0}\right).
\end{align}
For the variation of $S_B[U]$ we use
\begin{align}
\sum_{i,j} \dd_s \lb C_{x,ij} C^\dg_{x,ij} \rb & \sim \sum_{i, j} \left(\delta U_{x,i}U_{x+i,j}+U_{x,i}\delta U_{x+i,j} - \dd U_{x,j} U_{x+j,i} - U_{x,j} \dd U_{x+j,i}\right)C_{x,ij}^{\dg}+\hc \nonumber \\
& \sim i \sum_{i, j} 2 \delta A_{x,i}U_{x,i}\left(U_{x+i,j}C_{x,ij}^{\dg}+C_{x-j,ji}^{\dg}U_{x-j,j}\right)+\hc
\end{align}
The variation then reads
\bb
\dd_s S_B[U] = -\frac{V}{g^{2}}\sum_{x,i,j,a}\frac{1}{\left(a^{i}a^{j}\right)^{2}} \dd A_{x,i}^{a} P^{a}\left(U_{x,i}\left(U_{x+i,j}C_{x,ij}^{\dg}+C_{x-j,ji}^{\dg}U_{x-j,j}\right)\right).
\ee
We set $\dd S = 0$ and after canceling some constants we find the discrete EOM
\bb
\frac{1}{\left(a^{0}a^{i}\right)^{2}}P^{a}\left(U_{x,i0}+U_{x,i-0}\right)=-\sum_{j}\frac{1}{\left(a^{i}a^{j}\right)^{2}}P^{a}\left(U_{x,i}\left(U_{x+i,j}C_{x,ij}^{\dg}+C_{x-j,ji}^{\dg}U_{x-j,j}\right)\right),
\ee
which can also be written as
\bb
\frac{1}{\lb a^0 a^i \rb^2} P^a \lb U_{x,i0} + U_{x,i-0}\rb =
\sum_j \frac{1}{\lb a^i a^j \rb^2} P^a \lb U_{x,ij} + U_{x,i-j}\rb.
\ee
\section{Conservation of the Gauss constraint in the leapfrog scheme}\label{app_gauss}
We now explicitly show that the leapfrog EOM preserve the associated Gauss constraint. We use the identity for the fundamental representation of SU(N)
\bb
\sum_a t^{a}P^{a}\left(X\right) = \frac{1}{2i}\left(X-X^{\dg}\right)-\frac{1}{N}\tr\left(\frac{1}{2i}\left(X-X^{\dg}\right)\right) \one,
\ee
which can be shown using the Fierz identity for the generators of $\mathfrak su (N)$
\bb
\sum_a t_{ij}^{a}t_{kl}^{a}=\frac{1}{2}\left(\delta_{il}\delta_{jk}-\frac{1}{N}\delta_{ij}\delta_{kl}\right),
\ee
where $i,j,k,l$ are fundamental representation matrix indices. Using a shorthand we can write
\bb
\sum_a t^{a}P^{a}\left(X\right) = \ah{X},
\ee
where ``ah" denotes the anti-hermitian traceless part of $X$. The constraint and the equations of motion then read (including external charges)
\begin{align}
\sum_{i}\frac{1}{\left(a^{0}a^{i}\right)^{2}}\ah{U_{x,0i}+U_{x,0-i}} &= \frac{g}{a^0} \rho_x,\label{eq:app_gauss_ah} \\
\frac{1}{\lb a^0 a^i \rb^2} \ah{U_{x,i0} + U_{x,i-0}} &=
\sum_j \frac{1}{\lb a^i a^j \rb^2} \ah{U_{x,ij} + U_{x,i-j}} - \frac{g}{a^i} j_{x,i}. \label{eq:app_eom_ah}
\end{align}
We take \eqref{eq:app_eom_ah} and sum over $i$. Due to $\ah{U_{x,ij}}$ being antisymmetric in the index pair $i,j$ we find
\bb
\sum_i \frac{1}{\lb a^0 a^i \rb^2} \ah{U_{x,i0} + U_{x,i-0}} =
\sum_{i,j} \frac{1}{\lb a^i a^j \rb^2} \ah{U_{x,i-j}} - \sum_i \frac{g}{a^i} j_{x,i}.
\ee
Doing the same at $x-i$ and parallel transporting from $x-i$ to $x$ yields
\bb
\sum_i \frac{1}{\lb a^0 a^i \rb^2} \ah{U_{x,0-i} + U_{x,-0-i}} =
\sum_{i,j} \frac{1}{\lb a^i a^j \rb^2} \ah{U_{x,j-i}} - \sum_i \frac{g}{a^i} U^\dg_{x-i,i} j_{x-i,i} U_{x-i,i}.
\ee
Subtracting the above two equations gives
\bb
\sum_i \frac{1}{\lb a^0 a^i \rb^2} \ah{U_{x,i0} + U_{x,i-0} - U_{x,0-i} - U_{x,-0-i}} =
 - \sum_i \frac{g}{a^i} \lb j_{x,i} - U^\dg_{x-i,i} j_{x-i,i} U_{x-i,i} \rb.
\ee
Using antisymmetry we have 
\bb
\ah{U_{x,i0}} = - \ah{U_{x,0i}}
\ee
and
\bb
\ah{U_{x,-0-i}} = - \ah{U_{x,-i-0}}.
\ee
Moreover, in temporal gauge we have $U_{x,i-0} = U_{x-0,0i}$ and $U_{x,-i-0}=U_{x-0,0-i}$, which leads to
\begin{align}
\sum_i \frac{1}{\lb a^0 a^i \rb^2} \ah{U_{x,i0} + U_{x,i-0} - U_{x,0-i} - U_{x,-0-i}} &= \nonumber \\
\sum_i \frac{1}{\lb a^0 a^i \rb^2} \ah{-\lb U_{x,0i} + U_{x,0-i} \rb + \lb U_{x-0,0i} + U_{x-0,0-i} \rb} &= \nonumber \\
-\frac{g}{a^0} \lb \rho_x -\rho_{x-0} \rb, &
\end{align}
where we used the Gauss constraint in the last line to replace the temporal plaquette terms with charge densities.
This yields the gauge-covariant continuity equation
\bb
\frac{1}{a^0} \lb \rho_x -\rho_{x-0} \rb = \sum_i \frac{1}{a^i} \lb j_{x,i} - U^\dg_{x-i,i} j_{x-i,i} U_{x-i,i} \rb.
\ee
If there are no external charges, we simply have the conservation of the Gauss constraint: assume that the constraint in the previous time slice holds, i.e.
\bb
\sum_i \frac{1}{\lb a^0 a^i \rb^2} \ah{U_{x-0,0i} + U_{x-0,0-i}} = 0,
\ee
then the EOM guarantee that it will also hold in the next one, i.e.
\bb
\sum_i \frac{1}{\lb a^0 a^i \rb^2} \ah{U_{x,0i} + U_{x,0-i}} = 0.
\ee

\section{Variation of the implicit action}\label{app_implicit}

We now consider the action
\bb
S[U] = S_E[U] - S_B[U],
\ee
where
\begin{align}
S_E[U] &= \frac{V}{g^2} \sum_{x,i} \frac{1}{\lb a^0 a^i \rb^2} \tr \lb C_{x,0i} C^\dg_{x,0i} \rb,\\
S_B[U] &= \frac{V}{g^2} \sum_{x,i,\abs j} \frac{1}{4} \frac{1}{\lb a^i a^j \rb^2} \tr \lb C_{x,ij} M^\dg_{x,ij} \rb.
\end{align}
The electric part is the same as in the leapfrog scheme. However, the magnetic part $S_B[U]$ now contains temporal gauge links and gives a new contribution to the Gauss constraint. 

Before we vary this action, we must verify that $S_B[U]$ is indeed real-valued. While it is easy to see that the original leapfrog action is real-valued because of the obvious hermicity of $C_{x,ij} C^\dg_{x,ij}$, the term $C_{x,ij} M^\dg_{x,ij}$ is not hermitian in general. Still, we can show that the action is real: we have
\bb
C_{x,ij} M^\dg_{x,ij} = \frac{1}{2} C_{x,ij} \lb C^{(+0)}_{x,ij} + C^{(-0)}_{x,ij} \rb^\dg,
\ee
where we can rewrite
\begin{align}
C_{x,ij} C^{(+0)\dg}_{x,ij} &= C_{x,ij} \lb U_{x,0} C_{x+0,ij} U^\dg_{x+i+j,0}\rb^\dg \nonumber \\
&= C_{x,ij} U_{x+i+j,0} C^\dg_{x+0,ij} U^\dg_{x,0} \nonumber \\
&\sim U^\dg_{x,0} C_{x,ij} U_{x+i+j,0} C^\dg_{x+0,ij} \nonumber \\
&= C^{(-0)}_{x+0,ij} C^\dg_{x+0,ij}.
\end{align}
Here we used the cyclicity of the trace. Then using a shift $x \rightarrow x-0$ we have 
\bb
C_{x,ij} C^{(+0) \dg}_{x,ij} \sim C^{(-0)}_{x,ij} C^\dg_{x,ij}.
\ee
Likewise we have 
\bb
C_{x,ij} C^{(-0) \dg}_{x,ij} \sim C^{(+0)}_{x,ij} C^\dg_{x,ij},
\ee
which leads to
\bb
C_{x,ij} M^\dg_{x,ij} \sim M_{x,ij} C^\dg_{x,ij}.
\ee
Incidentally, the RHS term is exactly the hermitian conjugate of the LHS term. In other words
\bb
\lb C_{x,ij} M^\dg_{x,ij} \rb^\dg = M_{x,ij} C^\dg_{x,ij} \sim C_{x,ij} M^\dg_{x,ij}.
\ee
This means that under the sum over $x$ and the trace, the expression $C_{x,ij} M^\dg_{x,ij}$ is indeed real-valued and by extension $S_B[U]$ is real-valued as well. We have also shown that the time-average in $M_{x,ij}$ can be ``shifted" to the other term $C_{x,ij}$ under the sum and trace, i.e.
\bb
C_{x,ij} \avg{C}^\dg_{x,ij} \sim \avg{C}_{x,ij} C^\dg_{x,ij}.
\ee
This is a useful property that we will need in the following derivation.

\subsection{Gauss constraint} \label{app_implicit_gauss}
We perform the variation of $S_E[U]$ and $S_B[U]$ w.r.t.\ temporal links to derive the Gauss constraint in the implicit scheme. Since $S_E[U]$ is the same for all schemes, we do not have to repeat it. On the other hand, the variation of the magnetic part involves terms like $\dd_t \lb C_{x,ij} M^\dg_{x,ij} \rb$, which we now discuss explicitly. First, we make use of
\bb
\dd_t \lb C_{x,ij} M^\dg_{x,ij} \rb \sim \dd_t M_{x,ij} C^\dg_{x,ij} = \frac{1}{2} \lb \dd_t C^{(+0)}_{x,ij} + \dd_t C^{(-0)}_{x,ij} \rb C^\dg_{x,ij}.
\ee
Then, after some algebra we find
\bb
\dd_t C^{(+0)}_{x,ij} C^\dg_{x,ij} \sim i \dd A_{x,0} \lb C^{(+0)}_{x,ij} C^{\dg}_{x,ij} - C_{x,-i-j} \lb C^{(+0)}_{x,-i-j} \rb^\dg \rb
\ee
and 
\begin{align}
\dd_t C^{(-0)}_{x,ij} C^\dg_{x,ij} &\sim i \dd A_{x,0} \lb - C_{x,ij} \lb C^{(+0)}_{x,ij} \rb^\dg + C^{(+0)}_{x,-i-j} C^\dg_{x,-i-j} \rb \nonumber \\
&= \lb \dd_t C^{(+0)}_{x,ij} C^\dg_{x,ij} \rb^\dg,
\end{align}
which yields
\bb
\dd_t M_{x,ij} C^\dg_{x,ij} \sim \frac{i}{2} \dd A_{x,0} \lb \left[ C^{(+0)}_{x,ij} C^{\dg}_{x,ij} - \hc \right] + \left[  C^{(+0)}_{x,-i-j} C^\dg_{x,-i-j} -\hc \right] \rb.
\ee
The variation of the magnetic part therefore reads
\begin{align} \label{eq:var_imp_SB_t}
\dd_t S_B[U] =& - \frac{V}{g^2} \sum_{x,a,i,\abs j} \frac{1}{8} \frac{1}{\lb a^i a^j \rb^2} \dd A^a_{x,0} P^a \lb
C^{(+0)}_{x,ij} C^{\dg}_{x,ij} + C^{(+0)}_{x,-i-j} C^\dg_{x,-i-j}
\rb \nonumber \\
&= - \frac{V}{g^2} \sum_{x,a,\abs i,\abs j} \frac{1}{8} \frac{1}{\lb a^i a^j \rb^2} \dd A^a_{x,0} P^a\lb
C^{(+0)}_{x,ij} C^{\dg}_{x,ij} 
\rb.
\end{align}
In the last line we consolidated terms with index $i$ and $-i$ into a single term using the sum $\sum_{\abs i}$.
Taking the result for $\dd_t S_E[U]$ from the leapfrog scheme, we find the Gauss constraint
\bb
\sum_{i}\frac{1}{\lb a^{0}a^{i} \rb^{2}}P^{a}\lb U_{x,0i}+U_{x,0-i}\rb = 
- \sum_{\abs i,\abs j} \frac{1}{8} \frac{1}{\lb a^i a^j \rb^2} P^a\lb C^{(+0)}_{x,ij} C^{\dg}_{x,ij} \rb.
\ee
\subsection{Equations of motion} \label{app_implicit_eom}
For the EOM we vary $S[U]$ w.r.t\ spatial links. Again, we already have the result for $\dd_s S_E[U]$ from the leapfrog scheme and only need to calculate $\dd_s S_B[U]$. In particular we consider the term 
\bb
\dd_s \lb C_{x,ij} M^\dg_{x,ij} \rb = \dd_s  C_{x,ij} M^\dg_{x,ij} + C_{x,ij} \dd_s M^\dg_{x,ij}.
\ee
Since the variation only acts on spatial links we can shift the time-average of the right term from $M^\dg_{x,ij}$ to $C_{x,ij}$. This gives
\begin{align}
\dd_s \lb C_{x,ij} M^\dg_{x,ij} \rb &\sim \dd C_{x,ij} M^\dg_{x,ij} + M_{x,ij} \dd C^\dg_{x,ij} \nonumber \\
&= \dd C_{x,ij} M^\dg_{x,ij} + \hc
\end{align}
The variation then proceeds analogously to the derivation of the leapfrog scheme. We find
\bb \label{eq:var_imp_SB_s}
\dd_s S_B[U] =-\frac{V}{g^2} \sum_{x,i,\abs j} \frac{1}{2} \frac{1}{\lb a^{i}a^{j}\rb^{2}} \dd A_{x,i}^{a} P^{a}\lb U_{x,i}\lb U_{x+i,j} M_{x,ij}^{\dg}+M_{x-j,ji}^{\dg} U_{x-j,j}\rb \rb.
\ee
With the result for $\dd_s S_E[U]$ we obtain the discrete EOM
\bb
 \frac{1}{\lb a^0 a^i \rb^2} P^a \lb U_{x,i0} + U_{x,i-0}\rb =
- \frac{1}{2} \sum_{\abs j} \frac{1}{\lb a^i a^j \rb^2} P^a \lb U_{x,i} \lb U_{x+i,j} M^\dg_{x,ij} + M^\dg_{x-j,ji} U_{x-j,j}\rb\rb.
\ee
Introducing the shorthand
\bb
K_{x,ij}[U, M] = - \frac{1}{2} \frac{1}{\lb a^i a^j \rb^2} \lb U_{x+i,j} M^\dg_{x,ij} - M^\dg_{x-j,ij} U_{x-j,j}\rb,
\ee
allows us to write the EOM rather compactly as
\bb
 \frac{1}{\lb a^0 a^i \rb^2} P^a \lb U_{x,i0} + U_{x,i-0}\rb =
\sum_{\abs j} P^a \lb U_{x,i} K_{x,ij}[U,M] \rb.
\ee
\section{Variation of the semi-implicit action}\label{app_semi}

In the semi-implicit scheme the action reads
\bb
S[U] = S_E[U] - S_B[U],
\ee
where $S_E[U]$ is the same as before. The magnetic part comprises of $S_B[U] = S_{B,M}[U] + S_{B,W}[U]$, where $S_{B,M}[U]$ is the same as $S_B[U]$ from the implicit scheme except that $\sum_{i \abs j}$ only runs through transverse components:
\bb
S_{B,M}[U] = \frac{V}{g^2} \sum_{x,i,\abs j} \frac{1}{4} \frac{1}{\lb a^i a^j \rb^2} \tr \lb C_{x,ij} M^\dg_{x,ij} \rb.
\ee
Therefore we can take the results from the previous section for $\dd S_{B,M}[U]$, eqs.\ \eqref{eq:var_imp_SB_t} and \eqref{eq:var_imp_SB_s}. The new part is given by
\bb
S_{B,W}[U] = \frac{V}{g^2} \sum_{x, \abs j} \frac{1}{4} \frac{1}{\lb a^1 a^j \rb^2}  \tr \lb C_{x,1j} W^\dg_{x,1j} + \hc \rb.
\ee
\subsection{Gauss constraint} \label{app_semi_gauss}
We already know $\dd_t S_E[U]$ and $\dd_t S_{B,M}[U]$ from previous sections, so we only have to compute $\dd_t S_{B,W}[U]$. The relevant terms are
\begin{align}
\sum_{\abs j} \frac{1}{\lb a^j \rb^2}\dd_t W_{x,1j} C^\dg_{x,1j} &= \sum_{\abs j} \frac{1}{\lb a^j \rb^2} \lb \dd_t \avg{U}_{x,1} U_{x+1,j} - U_{x,j} \dd_t \avg{U}_{x+j,1} \rb C^\dg_{x,1j} \nonumber \\
&\sim \dd_t \avg{U}_{x,1} \sum_{\abs j} \frac{1}{\lb a^j \rb^2} \lb U_{x+1,j} C^\dg_{x,1j} - C^\dg_{x-j,1j} U_{x-j,j} \rb \nonumber \\
&= \dd_t \avg{U}_{x,1} T^{\dg}_{x,1},
\end{align}
where we defined
\bb
T^{\dg}_{x,1} = \sum_{\abs j} \frac{1}{\lb a^j \rb^2} \lb U_{x+1,j} C^\dg_{x,1j} - C^\dg_{x-j,1j} U_{x-j,j} \rb.
\ee
Using the same techniques as before we find 
\begin{align}
\dd_t \avg{U}_{x,1} T^{\dg}_{x,1} \sim \frac{i}{2} \dd A_{x,0} 
\bigg( & U^{(+0)}_{x,1} T^\dg_{x,1} - U_{x,1} T^{(+0) \dg}_{x,1} \nonumber \\
- &  T^\dg_{x-1,1} U^{(+0)}_{x-1,1} + T^{(+0) \dg}_{x-1,1} U_{x-1,1}  \bigg),
\end{align}
and the variation of $S_{B,W}[U]$ reads
\begin{align}
\dd_t S_{B,W}[U] = - \frac{V}{g^2} \sum_{x,a} \frac{1}{8\lb a^1 \rb^2} \dd A^a_{x,0} P^a \bigg(
&U^{(+0)}_{x,1} T^\dg_{x,1} + T^{(+0)}_{x,1} U^\dg_{x,1} \nonumber \\
+ &U^{(+0)\dg}_{x-1,1} T_{x-1,1} +  T^{(+0) \dg}_{x-1,1} U_{x-1,1},
\bigg)
\end{align}
which (with the previous results taken into account) yields the Gauss constraint in the semi-implicit scheme, see eq.\ \eqref{eq:semi_gauss}.

\subsection{Equations of motion} \label{app_semi_eom}

For the variation w.r.t.\ spatial links we have to distinguish two cases: the longitudinal and the transverse links. Starting with the variation of longitudinal links we find
\begin{align}
\dd_1 \lb C_{x,1j} W^\dg_{x,1j} \rb +\hc &=
\dd_1 C_{x,1j} W^\dg_{x,1j} + C_{x,1j} \dd_1 W^\dg_{x,1j} + \hc \nonumber \\
&\sim i \dd A_{x,1} U_{x,1} \bigg( \lb U_{x+1,j} W^\dg_{x,1j} + W^\dg_{x-j,j1} U_{x-j,j} \rb \nonumber \\
& + \avg{\lb U_{x+1,j} C^\dg_{x,1j} + C^\dg_{x-j,j1} U_{x-j,j} \rb}\bigg) + \hc
\end{align}
Again we used the fact that the time-average can be shifted to other terms by exploiting the sum over $x$ and the cyclicity of the trace. The variation of $S_{B,W}[U]$ then reads
\begin{align}
\dd_1 S_{B,W}[U] = -\frac{V}{g^2} \sum_{x,a,j} \frac{1}{4} \frac{1}{\lb a^1 a^j \rb^2} \dd A^a_{x,1} P^a \bigg(
U_{x,1} \bigg( &\lb U_{x+1,j} W^\dg_{x,1j} + W^\dg_{x-j,j1} U_{x-j,j} \rb \nonumber \\
+&\avg{\lb U_{x+1,j} C^\dg_{x,1j} + C^\dg_{x-j,j1} U_{x-j,j} \rb}\bigg) \bigg).
\end{align}
Combining the above with $\dd_1 S_E[U]$ gives the longitudinal EOM eq.\ \eqref{eq:semi_eom_1}. No contributions from $S_{B,M}[U]$ are necessary because it does not include any longitudinal links.

For the transverse components of the EOM we vary w.r.t.\ $U_{x,j}$, where $j$ is a transverse index. The relevant terms for $j>0$ are
%
%
%
\begin{align}
\dd_j W_{x,1j} C^\dg_{x,1j} \sim i \dd A_{x,j} U_{x,j} \lb \avg{U}_{x+j,1} C^\dg_{x,j1} - C^\dg_{x-1,j1} \avg{U}_{x-1,1} \rb,
\end{align}
and
\begin{align}
\dd_j C_{x,1j} W^\dg_{x,1j} \sim i \dd A_{x,j} U_{x,j} \lb U_{x+j,1} W^\dg_{x,j1} - W^\dg_{x-1,j1} U_{x-1,1} \rb.
\end{align}
For terms with negative component indices $j<0$ we can show that they are identical (under the sum over $x$ and the trace) to the last two terms except for the substitution $1 \rightarrow -1$.
\begin{align}
\dd_j \sum_{-j}  \lb W_{x,1j} C^\dg_{x,1j} + \hc \rb &= \dd_j \sum_{j} \lb W_{x,1-j} C^\dg_{x,1-j} + \hc \rb \nonumber \\
&\sim \sum_{j} \lb \dd_j W_{x,-1j} C^\dg_{x,-1j}  + \dd_j C_{x,-1j} W^\dg_{x,-1j} + \hc \rb.
\end{align}
This allows us to write
\begin{align}
\sum_{\abs j} \dd_j W_{x,1j} C^\dg_{x,1j} &\sim \sum_{\abs 1} i \dd A_{x,j} U_{x,j} \lb \avg{U}_{x+j,1} C^\dg_{x,j1} - C^\dg_{x-1,j1} \avg{U}_{x-1,1} \rb, \\
\sum_{\abs j} \dd_j C_{x,1j} W^\dg_{x,1j} &\sim \sum_{\abs 1} i \dd A_{x,j} U_{x,j} \lb U_{x+j,1} W^\dg_{x,j1} - W^\dg_{x-1,j1} U_{x-1,1} \rb,
\end{align}
where $\sum_{\abs 1}$ stands for summing over terms with component indices $1$ and $-1$.
The variation of $S_{B,W}[U]$ then reads
\begin{align}
\dd_j S_{B,W}[U] =& \frac{V}{g^2} \sum_{x, \abs{j}} \frac{1}{4} \frac{1}{\lb a^1 a^j \rb^2}  \dd_j \tr \lb C_{x,1j} W^\dg_{x,1j} + \hc \rb & \nonumber \\
=& - \frac{V}{g^2} \sum_{x,j,a} \frac{1}{4} \frac{1}{\lb a^1 a^j \rb^2} \dd A^a_{x,j} P^a \bigg( U_{x,j} \sum_{\abs 1}\bigg( \lb \avg{U}_{x+j,1} C^\dg_{x,j1} - C^\dg_{x-1,j1} \avg{U}_{x-1,1} \rb \nonumber \\
& \qquad\qquad+ \lb U_{x+j,1} W^\dg_{x,j1} - W^\dg_{x-1,j1} U_{x-1,1} \rb \bigg) \bigg).
\end{align}
With the expressions for $\dd_j S_E[U]$ and $\dd_j S_{W,M}[U]$ we find the transverse components of the EOM eq.\ \eqref{eq:semi_eom_i}. 

\bibliographystyle{JHEP}
\bibliography{references}
\end{document}